\documentclass[oldversion]{aa}  
\usepackage{graphicx}
\usepackage{subfigure}
\usepackage{txfonts}
\usepackage{natbib}
\usepackage{rotating}
\usepackage{longtable,lscape}
\bibpunct{(}{)}{;}{a}{}{,}
\begin{document}
 \title{Low temperature FIR and submm mass absorption coefficient of interstellar silicate dust analogues}
 
\author{A. Coupeaud\inst{1,2}
 \and    K. Demyk\inst{1,2}
 \and    C.  Meny\inst{1,2}
 \and    C. Nayral\inst{3}
 \and    F. Delpech\inst{3}
 \and    H. Leroux\inst{4}
 \and    C. Depecker\inst{4}
 \and    G. Creff \inst{5}
 \and    J.-B. Brubach\inst{5}
 \and    P. Roy\inst{5}
 }

  \offprints{demyk@cesr.fr}

\institute{Universit\'e de Toulouse; UPS-OMP; IRAP;  Toulouse, France
\and
CNRS; IRAP; 9 Av. colonel Roche, BP 44346, F-31028 Toulouse cedex 4, France
\and
Universit\'e de Toulouse, INSA, CNRS, LPCNO, F-31077 Toulouse, France
\and
UMET, UMR 8207, Universit\'e Lille 1, CNRS, F-59655 Villeneuve d'Ascq, France
\and
Ligne AILES - Synchrotron SOLEIL, L\'\ Orme des Merisiers, F-91192 Gif-sur-Yvette, France }

   \date{Received 23 March 2011; accepted 07 September 2011}

  \abstract
   { Cold dust grains are responsible for the far infrared and submillimetre (FIR/submm) emission observed by Herschel and Planck. Their thermal emission is usually expressed as a modified black body law in which the FIR/submm dust opacity, or mass absorption coefficient (MAC), is described by the  mass absorption coefficient at a given wavelength, ${\kappa}_{\lambda_0}$, and the temperature- and wavelength-independent emissivity spectral index, $\beta$. However, numerous data from previous space and balloon-born missions and recently from Herschel and Planck show that the dust emission is not well understood, as revealed for example by the observed anti-correlation of $\beta$ with the grain temperature. \\The aim of this work is to measure the optical properties of interstellar dust analogues at low temperatures to give astronomers the necessary data to interpret FIR/submm observations such as those from the Herschel and Planck satellites.\\ We synthesised, via sol-gel methods, analogues of interstellar amorphous silicate grains, rich in Mg and Ca, and having stiochiometry of olivine and pyroxene. The samples are characterised by various techniques to determine their composition, size, amorphization degree. All the amorphous samples are annealed at 1100$^{\circ}$C to study the crystallised materials for comparison. We measured the MAC of all the samples, in the 2-25 $\mu$m range at room temperature and in the 100-1000/1500 $\mu$m range for grain temperatures varying from 300 to 10 K. \\ The experimental results show that for all the amorphous samples, the grain MAC decreases when the grain temperature decreases and that the local spectral index, $\beta$, defined as the slope of the MAC curve, is anti-correlated with the grain temperature. These variations, which are not observed in the crystallised samples, are related to the amorphous nature of the samples. In addition, the spectral shape of the MAC is complex: it cannot be described by a single spectral index over the 100-1500 $\mu$m range. At short wavelengths ($\lambda \le$ 500/700 $\mu$m), $\beta$ is in the range 1.6 - 2.1 for all grain temperature and grain composition. However, at longer wavelengths  ($\lambda \ge$ 500/700 $\mu$m), $\beta \le $ 2 for samples with a pyroxene stoichiometry and $\beta \ge $ 2 for samples with an olivine stoichiometry. \\ The dust properties in the FIR/submm domain and at low temperature are more complicated than expected. The simplifying asymptotic expression based on a single temperature- and wavelength-independent spectral index used by astronomers is not appropriate to describe the dust MAC and hence the dust emission, and may induce significant errors on the derived parameters such as the dust mass and the dust physical and chemical properties. Instead, dust emission models should use the dust MAC as a function of  wavelength and temperature. }


\maketitle

\section{Introduction}
\label{intro}
\noindent The far infrared and submillimetre emission of the Interstellar Medium (ISM) is dominated by grains of size of the order of a few tens to a few hundreds of nanometers, mostly composed of silicates and oxides \citep{desert1990,compiegne2011}. These grains are ubiquitous. Their physico-chemical properties and their evolution within the lifecycle of the ISM are constrained by the spectroscopic observations in the mid infrared spectral range (2-50 $\mu$m) via the study of the vibration of the chemical bonds constituting the grains. Results from the space-born satellites ISO (Infrared Space Observatory) and Spitzer show that although crystalline silicates such as forsterite (Mg$_2$SiO$_4$) and enstatite (MgSiO$_3$) are observed around stars, the silicates in the ISM are completely amorphous \citep{kemper2004}. It is more difficult to study amorphous silicates than crystalline silicates because the vibrational bands at $\sim$ 9.7 $\mu$m and $\sim$ 18 $\mu$m of amorphous silicates are broad and structureless whereas in crystalline silicates, they exhibit numerous sharp spectral features due to the long-range order of the crystalline lattice. Consequently, the composition of the amorphous  silicates is not as well constrained as for crystalline silicates. Nevertheless, the comparison of astrophysical spectra with laboratory spectra of amorphous silicate dust analogues and the constraints coming from the depletion of the elements from the gas-phase, indicate that the amorphous silicates should have an average composition close to the one of the crystalline silicates: Mg$_2$SiO$_4$, MgSiO$_3$. In addition dust grains certainly contain iron, which is highly depleted, but the form in which it is incorporated to the grains is unclear (within the silicate structure, in oxides, in the form of iron inclusions). They may also contain other minor elements highly depleted from the gas-phase such as calcium or aluminium. These elements may be incorporated within the silicates chemical network or in the form of oxides. 

\noindent Observations in the far infrared and submillimetre range trace cold dust (T$\sim$10-40 K). It allows us to detect pre-stellar cores, prior to their collapse and thus to study the very first steps of star formation. It is used to estimate the dust mass and density of interstellar cold clouds. When the interstellar grains are big enough (above 10 nm), the absorption of UV-photons from the ambient radiation field does not induce a significant grain temperature variation and these cold grains radiate at thermal equilibrium. In cold environments of the ISM, this emission dominates above other dust component emission at wavelength greater than 100 $\mu$m. Assuming an optically thin medium, which is usually the case in the FIR/mm wavelength range, their modified blackbody emission is expressed as:
\begin{equation} \label{emission}
\mathrm{I(\lambda) = B(\lambda,T)\:\tau(\lambda)  = B(\lambda,T)\: \kappa(\lambda) \: M_{d} \: d^{-2} \: \Omega^{-1}}
\end{equation}  
 where I$_{\lambda}$ is the specific intensity and B$_{\lambda}$(T) the Planck function at the dust temperature in W.m$^{-2}$.Hz$^{-1}$.sr$^{-1}$, M$_{d}$ the mass of dust along the line of sight, d the distance of the source, $\Omega$ the beam solid angle and $ \kappa(\lambda)$ the mass absorption coefficient (MAC, cm$^2$.g$^{-1}$). Simple semi-classical physical models of absorption, provide, at long wavelength, a temperature independent asymptotic behaviour for the dust emissivity that is expressed as follow:  
 \begin{equation} \label{kappa}
\mathrm{\kappa =\:\kappa _{\lambda_{0} }\: ( \frac{\lambda}{\lambda}_{0}) ^{-\beta}}
\end{equation}  
where the emissivity spectral index, $\beta$, is ranging from 1 to 2, a value of 2 being commonly adopted in astrophysical studies. 

\noindent Before the launch of the Herschel and Planck satellites the, relatively few, observations in the far infrared and submillimetre wavelength revealed that the dust emission was not easy to interpret. Among the observed anomalous behaviours are the far-infrared excess observed by the instrument FIRAS on board the COBE (Cosmic Background Explorer) satellite (Reach et al. 1995) and the $\beta$-T anti-correlation derived from the observations of the two balloon-born missions PRONAOS and ARCHEOPS \citep{dupac2003,desert2008}. The FIRAS observations revealed the existence of a significant millimetre excess in the dust emission spectrum, with respect to a modified black body law with a spectral index $\beta$ equals to 2. The data can be well-fitted with a spectral index value of 1 instead of 2. It was proposed that this excess is due to the presence of very cold dust ($\sim$ 9 K), in addition to a warmer component responsible for the emission maximum near 100 $\mu$m \citep{finkbeiner1999}. However, the fact that the excess is well correlated with the FIR dust emission indicates that it could originate from unidentified processes intrinsic to the grains radiating in thermal equilibrium, without requiring an additional colder component. The deviation of the spectral index value from the classical value of 2 was also observed in various interstellar regions for which the $\beta$ value increases (up to 3 or 4) when the mean dust temperature decreases (down to 10 K) whereas it is as low as 1 when the dust temperature equals 80 K \citep{dupac2003,desert2008}. The anti-correlation of $\beta$ and T is now confirmed by the first results of the Herschel mission which show that the value of the spectral index, derived from the fit of the observed spectral energy distribution (SED) of regions in our Galaxy and in external galaxies, often deviates from the value of 2 \citep{anderson2010,rodon2010,bot2010}.  
The Planck Early Results confirm the dust emissivity variations and indicate that they are ubiquitous. First results show that the dust emissivity index in the solar neighbourhood of our Galaxy changes from $\beta$=1.8 in the FIR to a flatter behaviour in the submillimetre  \citep{Planck2011o}. They also clearly show that dust in nearby galaxies (the Large and Small Magellanic Clouds, LMC and SMC in particular) has a  significantly flatter FIR emissivity ($\beta$=1.5 and 1.2 respectively) than our Galaxy  \citep{Planck2011m}. Complementarily, using the new Herschel data, \cite{paradis2010} showed that the emissivity index of dust in our Galaxy actually varies with temperature in the far infrared. With its full sky survey and unprecedented sensitivity, Planck has also allowed us to detect more than 10\,000 cold clumps in the Galaxy \citep{Planck2011s}. The temperature of the Planck cold clumps spans from 7 K to 17 K, with $\beta$ ranging from 2.8 to 1.4. It has been shown that the data are not consistent with a constant value of the dust spectral index $\beta$ over the whole range of temperature, confirming the anti-correlated behaviour observed previously. Furthermore, Monte-Carlo simulations have demonstrated that such an anti-correlation cannot be explained by a fitting degeneracy given the error bars, and could be a real feature of dust emission \citep{desert2008,veneziani2010,paradis2010}. 

\noindent  Motivated by the results from infrared satellites, many experimental and theoretical studies have been undertaken to characterise the optical properties of silicate dust analogues in this spectral range. Most of the experimental studies focus on the MIR range (\citet{jaeger1994} and associated papers, \citet{koike2000}) although in some cases, experimental spectra were recorded up to 100 or 300 $\mu$m and sometimes at low temperatures \citep{henning1997,chihara2001,koike2006,mutschke2008}. However only a few studies were dedicated to the far infrared and submillimetre spectral range. In the far infrared range, the mass absorption coefficient (MAC) may be influenced by various factors such as the dust aggregation, the grain size, the presence of conducting particles (see \citet{stognienko1995,henning1995} for a modelisation of these effects). The  dust temperature may also play an important role as demonstrated by the experimental studies from \citet{agladze1996,mennella1998,boudet2005}. \cite{agladze1996} measured the absorption spectra of crystalline enstatite and forsterite and of amorphous Mg-rich submicronic silicate grains at very long wavelength (700~$\mu$m -2.9 mm) and very low temperature (1.2-30~K). \cite{mennella1998} performed similar measurements in the 20-2000 $\mu$m range on forsterite (Mg$_2$SiO$_4$) and amorphous and crystalline fayalite (Fe$_2$SiO$_4$) at higher temperatures (24-300~K). \cite{boudet2005} studied amorphous MgSiO$_3$ and silica (SiO$_2$) submicronic grains in the 100-2000 $\mu$m range and low temperature (10-300~K). These three studies all show that the absorption coefficients of the amorphous samples are frequency-dependent and temperature-dependent. However, because these studies were made on different materials, synthesised with different techniques and characterised at different temperatures and spectral ranges, they are difficult to compare. 

\noindent We present in this paper an experimental study of  the optical properties of interstellar silicate dust analogues in the wavelength range 100-1500 $\mu$m and from 300 to 10~K. The studied samples have a mean composition close to Mg$_2$SiO$_4$, MgSiO$_3$ and CaMgSi$_2$O$_6$. We study amorphous and crystallised samples to investigate the influence of disorder on the variation of the optical properties of the material as a function of the grain temperature. The synthesis of the analogues and their characterisation are presented in Sect.~\ref{sample}. The experimental details related to the spectroscopic measurements are described in Sect.~\ref{spectro}. The results are presented in Sect.~\ref{results} and discussed in Sect.~\ref{discussion}. The astrophysical implications are discussed in Sect.~\ref{astro}.

\begin{figure*}[!t]
\centering
   \includegraphics[scale=.241]{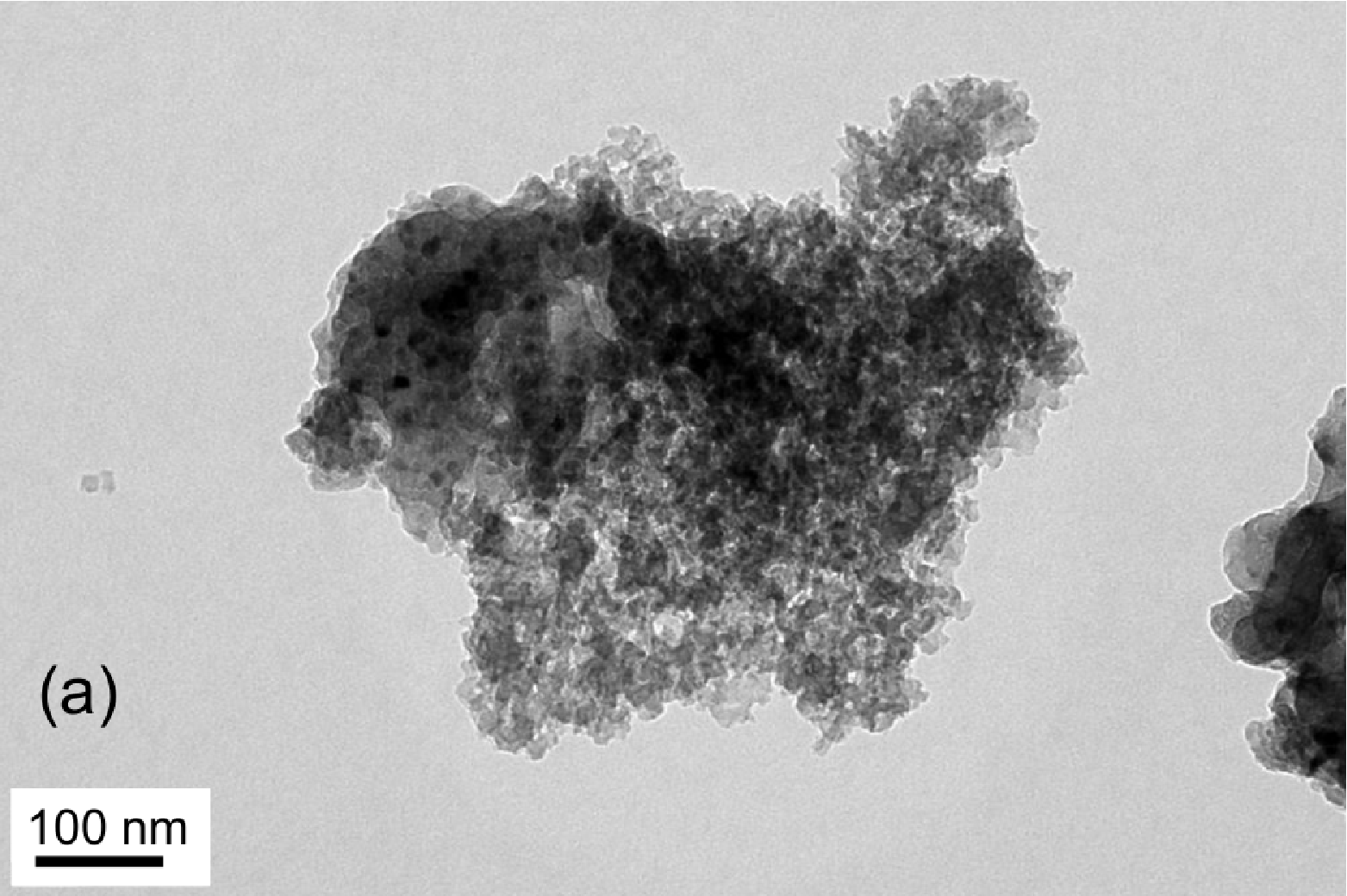}
   \includegraphics[scale=.205]{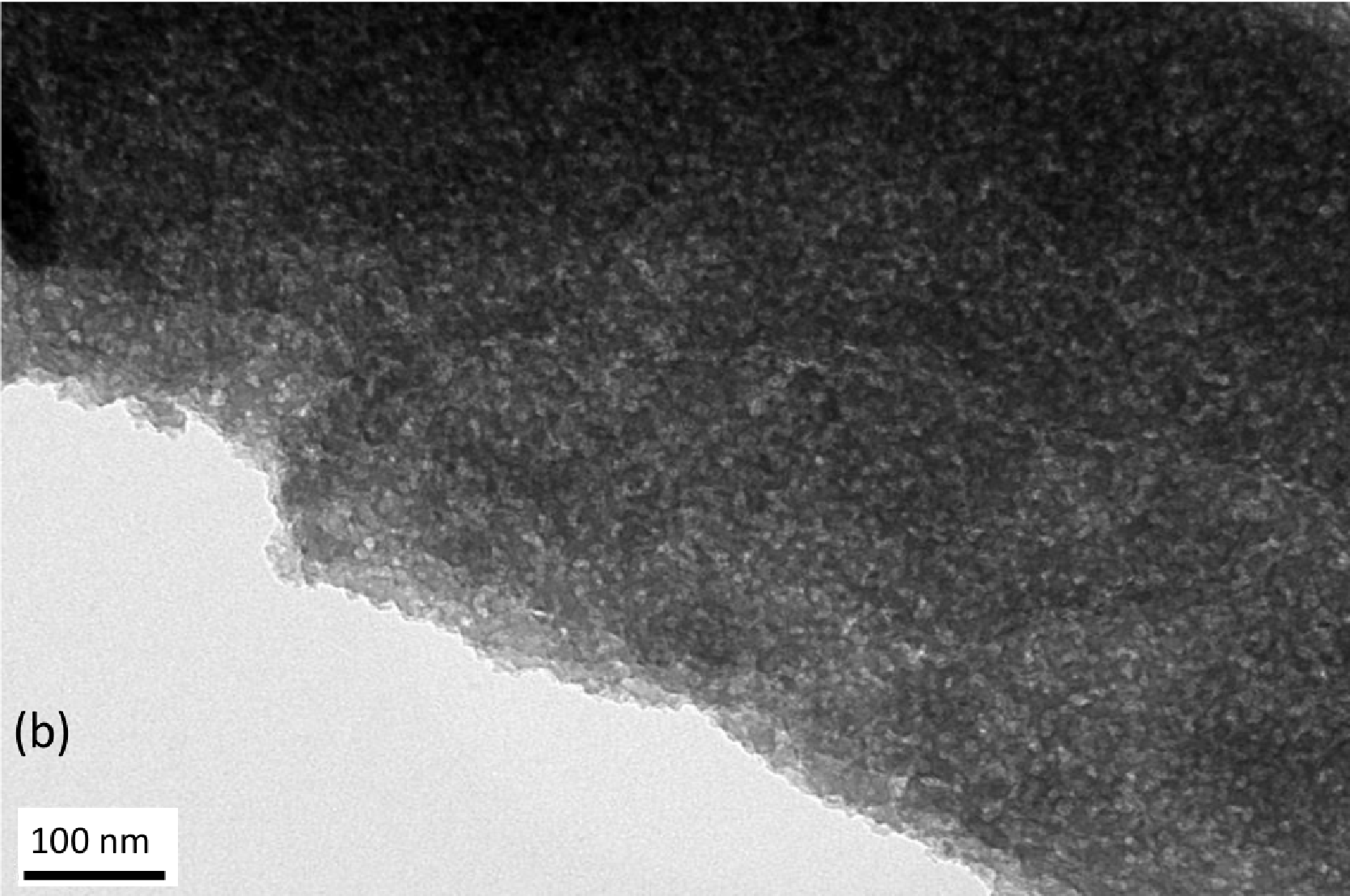}
   \includegraphics[scale=.575]{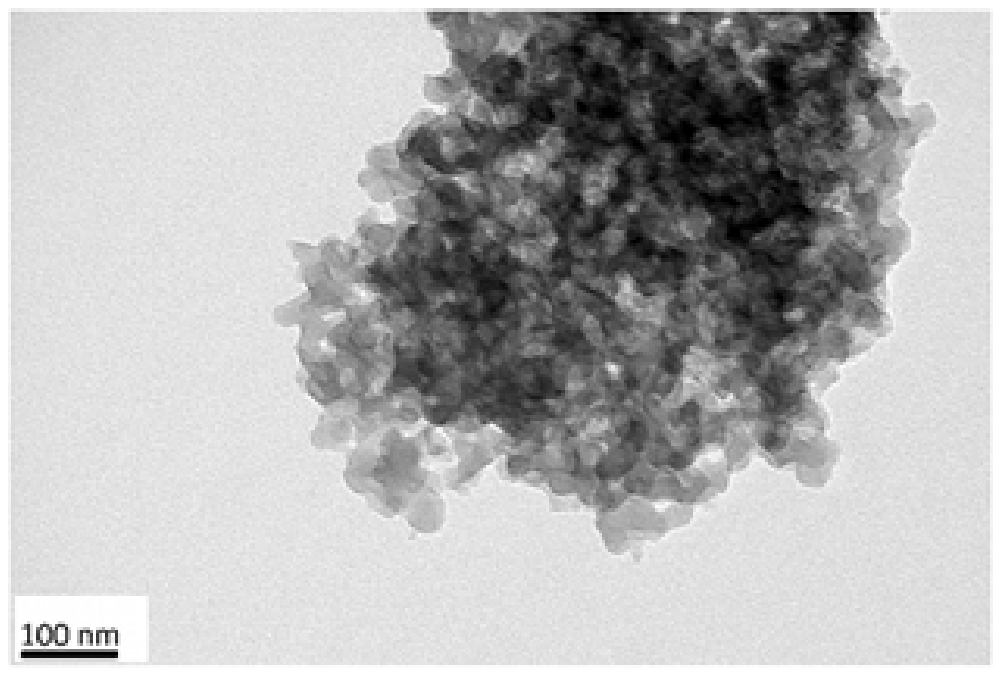}
          \caption{TEM bright field images of (a) sample F1, (b) sample F2 and (c) sample E. All samples are prepared by a sol gel method. The microstructure consist of small clusters stick together. The sample have a high porosity and a high surface/volume ratio. }
    \label{lte}
\end{figure*}

\begin{table*}
\caption{Characteristics of the synthesised dust analogues.} 
\label{samples}
\begin{center}
\begin{tabular}{ c c c c c c c  }
\hline \hline 
sample 	& sol-gel synthesis	& composition	 				& specific area & $ \phi$ pores \\ 
		&	recipe	   	&		 		& (m$^{2}$.g)	&  (nm)			      \\
\hline 
F1 & TEOS + Mg(OEt)$ _{2}$ + H$ _{2}$O + HCl 					   &  Mg$ _{2.3}$SiO$ _{4}$    	& -  & $\sim$ 5-10\tablefootmark{(1)} \\
F2 & TEOS + Mg(OEt)$ _{2}$ + H$ _{2}$O + C$ _{4}$H$ _{9}$NH$ _{2}$ &  Mg$ _{2.8}$SiO$ _{4}$    	& - & $\sim$ 5-10\tablefootmark{(1)}\\  
F3 & TEOS + Mg(NO$ _{3}$)$ _{2}$ + H$ _{2}$O + HNO$ _{3}$  &  Mg$ _{2.05}$SiO$ _{4}$ 	& - & $\sim$ 5-10\tablefootmark{(1)} \\ 
E & TEOS + Mg(NO$ _{3}$)$ _{2}$ + H$ _{2}$O  + HNO$ _{3}$   &  Mg$ _{0.95}$SiO$ _{3}$ 	& 132 & 9\tablefootmark{(2)}\\
D & TEOS  + Mg(NO$ _{3}$)$ _{2}$ + Ca(NO$ _{3}$)$ _{2}$ + H$ _{2}$O + HNO$ _{3}$ &  Ca$ _{0.98}$Mg$ _{0.9}$Si$ _{2}$O$ _{6}$ & 83 &  17\tablefootmark{(2)} \\  
\hline 
\hline 
\end{tabular} 
\tablefoot{\tablefoottext{1}{Estimated from TEM images.} 
\tablefoottext{2}{Derived from BET measurements. \\}
}\end{center}
\end{table*}

\section{Sample synthesis and characterisation}
\label{sample}

\subsection{Sample synthesis}

We have chosen to study different amorphous Mg-rich silicates representative of astronomical silicate dust (Table~\ref{samples}): 3 samples with a stoichoimetry close to olivine (Mg$_{2}$SiO$_{4}$) and 2 samples with a stoichiometry close to pyroxene ((Mg,Ca)$_2$Si$_2$O$_{6}$). The 3 silicates with an olivine stoichiometry were synthesised by sol-gel method using three different procedures. The sol-gel process is based on the hydrolysis and condensation of tetraethoxysilane (TEOS) in a solvent using water and a Mg-rich compound as the reactant, in presence of a catalyst. The main differences between the 3 adopted procedures are the choice of the Mg-rich compound, which can be a nitrate salt (Mg(NO$_{3}$)$_{2}$) or an alkoxide (Mg(OEt)$_{2}$), and the nature of the catalyst which can be acidic or basic. The first sample, called F1, was synthesised following the procedure of \citet{mitchell1998} using Mg(OEt)$_{2}$ and HCl. The second sample, called F2, was synthesised following a procedure from Coupeaud (unpublished) using Mg(OEt)$_{2}$ and butylamine (CH$_{3}$(CH$_{2})_{3}$NH$_{2}$, hereafter BA). For this method, a solution of TEOS (3~mmol, 0.12~M) in tetrahydrofurane and 2-methoxyethanol was prepared at room temperature with a TEOS:BA:H$_{2}$O ratio of 1:0.5:4 \citep{elhawi2009}. A solution of Mg(OEt)$_{2}$ (6~mmol, 0.24~M) prepared as described by \citet{mitchell1998} was then added to the TEOS solution and the mixture was stirred and heated at 50$^{\circ}$C for 12 days. Finally, the obtained suspension was centrifuged and washed with tetrahydrofurane and 2-methoxyethanol. The third sample, called F3, was synthesised following the procedure of \citet{gillot2009} using Mg(NO$_{3}$)$_{2}$ and HNO$_{3}$. The samples with a pyroxene composition, called E and D, were synthesised using the sol-gel method from \citet{gillot2009} using TEOS and Mg(NO$_{3}$)$_{2}$ with a ratio of 1 for the E sample and using TEOS, Mg(NO$_{3}$)$_{2}$ and Ca(NO$_{3}$)$_{2}$ for the D sample. To remove the organics from the synthesis, all the samples were heated up to 400$^{\circ}$C or 500$^{\circ}$C depending on the synthesis procedure. The composition of the synthesised samples were determined by Inductively Coupled Plasma Mass Spectroscopy (ICP-MS, Antellis Company), they are indicated Table~\ref{samples}.  All the samples contains OH groups and water coming from the sol-gel method that partly remain within the samples after thermal annealing at 400-500$^{\circ}$C. The influence of OH groups and water on the FIR optical properties of the samples is discussed in Sect.~\ref{discussion}. In addition, about half of each of theses amorphous samples was annealed for 24 hours at 1100$^{\circ}$C in oven under ambient atmosphere. The resulting samples, called F1c, F2c, F3c, Ec, and Dc, are crystalline as shown by the XRD and NMR analysis and MIR spectroscopy (see Sect.~\ref{xrd}, Sect.~\ref{nmr} and Sect.~\ref{s-mir}). Thus, a total of 10 samples are considered in this study: the 5 amorphous samples produced by sol-gel method and annealed at 400-500$^{\circ}$C and the 5 crystallised samples obtained by annealing of the amorphous samples at 1100$^{\circ}$C.

\subsection{Specific area and porosity}
\label{BET}
The surface area and average pore size of the E and D samples were determined by physical gas adsorption technique. An inert gas, nitrogen, is adsorbed on the surface of a solid material (on the outer surface and, in case of porous materials, also on the surface of the pores) at a temperature of 77 K leading to a so-called adsorption isotherm. It is standard to apply Brunauer-Emmett-Teller (BET) method to derive the surface area and pore size from the adsorption isotherm. The results show that the synthesised sol-gel silicates E and D are porous, with an average pores diameter between 10-20 nm,  and present a high specific area in the range 80 - 130 m$ ^{2} $/g. For the F1, F2 and F3 samples, no BET measurements were performed. 

\subsection{Transmission electron microscopy}
\label{s-TEM}
Transmission electron microscopy (TEM) is used to investigate the samples prepared by slow evaporation of a drop of the colloidal solution deposited onto a carbon covered copper grid. The TEM analysis were performed at the "Service Commun de Microscopie Electronique de l'Universit\'{e} Paul Sabatier" (TEMSCAN) on a JEOL JEM 1011 electron microscope operating at 100~kV with a point resolution of 4.5~{\AA}  and at the "Centre Commun de Microscopie de Lille" on a FEI Tecnai G2-20 twin microscope operating at 200~kV. The samples E and D (having a pyroxene stoichiometry) have an homogeneous composition. The TEM images of these two samples show that the particles present a complex texture with a fine apparent porosity with pore sizes of 5-20 nm, compatible with the BET analysis (see Sect.~\ref{BET}). The 3 samples having an olivine composition are different from each other. The F1 sample is chemically inhomogeneous: Mg-rich nanocrystal of $\sim$ 5 nm (presumably MgO) are identified on the TEM images together with regions deficient in magnesium (Mg/Si $\sim$ 1). The F2 sample is relatively stoichiometric with a slight Mg excess compare to Si. The F3 sample is partly crystallised. As for the samples with a pyroxene stoichiometry, samples with an olivine stoichiometry are characterised by a fine porosity with an apparent pore size of $\sim$ 5-10 nm. However no BET measurements were performed on the F samples to confirm this size.

\
\begin{table}[!t]
\caption {Composition of the crystalline phases measured by XRD after annealing at 1100$^{\circ}$C of the silicate samples.}
\label{t_xrd}
\begin{center}
\begin{tabular}{llccccc}
\hline \hline 
phase		& phase				          & F1c& F2c & F3c & Ec & Dc   \\ 
name		& composition				 &   \multicolumn{5}{c}{phase abundance (in mass \%)}	 \\ 
\hline 
periclase		& MgO  					  &  38	&  18  	&  6		&     -	&  -  	\\
forsterite		& Mg$_{2}$SiO$_{4}$ 		  &  38	&  82	&  94 	& 15 	& -   	\\  
orthoenstatite	& MgSiO$_{3}$  		  	  & 	24    & 	-	&	- 	& 64 	&  -   \\ 
clinoenstatite	& MgSiO$_{3}$ 	  	 	  & 	-     	& 	-	& 	-	& 21 	&  -   \\
merwinite		& Ca$_{3}$MgSi$_{2}$O$_{8}$ & 	-    	& 	-	& 	-	& 	-	& 7	\\  
diopside		& CaMgSi$_{2}$O$_{6}$ 		  &    -   	& 	-	& 	-	& 	-	& 65 \\
akermanite	& Ca$_{2}$MgSi$_{2}$O$_{7}$ &    -	& 	-	& 	-	& 	-	& 28 \\ 
\hline
\end{tabular} 
\end{center}
\end{table}

\subsection{X-rays diffraction}
\label{xrd}

X-rays diffraction (XRD) measurements are used to monitor the amorphous degree of the samples and to characterise the composition of the crystalline phases formed after annealing. The samples were ground in an agate mortar and were analysed in the Panalytical MPD Pro powder diffractometer at the Laboratoire de Chimie de Coordination (CNRS, Toulouse)  and in a Seifert XRD 3000TT diffractometer at the Centre d'Elaboration de Mat\'{e}riaux et d'Etudes Structurales (CNRS, Toulouse). The X-ray diffraction measurements were carried out in $\theta-\theta$ configuration and diffraction profiles were obtained using a step interval of respectively 0.017~$^{\circ}$ and 0.015~$^{\circ}$ in 2$\theta $ according to the diffractometer used. The analysis show that the samples annealed at 400 or 500$^{\circ}$C are amorphous, except for the F3 sample which XRD spectrum exhibits peaks characteristic of forsterite. This is confirmed by the NMR measurements and the MIR spectrum (see Sections \ref{nmr} and \ref{s-mir}).   \\
The quantitative analysis of the phases constituting each crystallised samples was performed by Rietveld refinement of the powder diffraction patterns \citep{bish1988}. The identification of the phases constituting each compound was performed using the Eva software ({\it http://www.bruker-axs.de/eva.html}). The identified phases were then used as an input for Rietveld refinement, using the software MAUD ({\it http://www.ing.unitn.it/~maud/}). This technique allows us to derive the relative concentration of the different crystalline phases with an error of about 5\%.
The Rietveld refinement of the F1c, F2c, F3c samples XRD measurements shows the existence of two crystalline phases, periclase (MgO) and forsterite (Mg$_2$SiO$_4$), respectively in the following ratios : 38\%/38\%, 18\%/82\% and 6\%/94\% (Table~\ref{t_xrd}). In addition,  around 24\% of orthoenstatite (MgSiO$_3$) is present in the F1c sample. The powder pattern of the sample Ec shows that the sample is a mixture of forsterite (15\%), orthoenstatite (64 \%) and clinoenstatite (21\%). The powder pattern of the sample Dc shows that the sample is a mixture of 3 crystalline phases: merwinite (7\%), diopside (65\%) and akermanite (28\%). 

\subsection{Nuclear magnetic resonance}
\label{nmr}

Solid-state NMR spectra of $^{29}$Si were recorded at the Laboratoire de Chimie de Coordination (CNRS, Toulouse) on a Bruker Avance 400WB spectrometer equipped with a 9,4~T magnet. NMR spectra are referenced to TMS. For all experiments we used a 4~mm diameter probe and a magic-angle spinning (MAS) rate of 12~kHz. MAS $^{29}$Si NMR spectroscopy using high power decoupling (HPDEC) and $^1$H-$^{29}$Si cross polarisation (CP) sequences is a powerful tool to characterise the silicate structure at the atomic scale. The chemical shift data of $^{29}$Si NMR spectra of silicate samples allow the identification of the different Q$^{n} $-type silicon signals (n represents the number of siloxy groups (-OSi-) per silicon atoms, Si-O-Si). Thus, Q$ ^{4}$ corresponds to Si(OSi)$_{4}$, Q$ ^{3}$ to -Si(OSi)$_{3}$, Q$ ^{2}$ to =Si(OSi)$_{2}$, Q$ ^{1}$ to $  \equiv$Si(OSi) and Q$ ^{0}$ to isolated tetrahedron SiO$_{4}$. \\ The analysis of crystalline samples is compatible with the XRD results. The NMR spectra exhibit narrow peaks with chemical shifts attributed to Q$ ^{0}$ and Q$^{2}$ groups that are characteristic of respectively forsterite and pyroxene (enstatite and diopside). \\ In amorphous samples, these peaks are broader than in crystals. In addition, because the connectivity in amorphous and crystalline materials is different, the amorphous samples present a wide distribution of Q$^{n} $ groups. The proportions of the Q$^{n} $ species were obtained from the deconvolution of the $^{29}$Si HPDEC NMR spectrum with the Dmfit program \citep{massiot2002}. The relative amount of Q$^{n} $ groups are correlated to the structure of the sample. Indeed, the F samples are characterised by a majority of Q$^{0} $ and Q$^{1} $ moieties ($\sim$60\%) and exhibit the features of Q$ ^{2}$, Q$ ^{3}$ and Q$ ^{4}$ too. In addition, the spectrum of F3 shows a spectral feature of Q$^{0} $ group which is very narrow and shifted in comparison with the amorphous Q$ ^{0}$ type silicon signal confirming that the sample is partially crystalline and contains $\sim$14\%  of forsterite. The deconvolution of the spectrum of the E sample reveals a majority of Q$^{1} $ and Q$^{2}$ groups ($ \backsim 70\% $) and the presence of Q$^{3} $ and Q$^{4} $ groups. For the D sample, the distribution of Q$^{n}$ species is made up of $ \backsim 40\% $ of  Q$^{2} $ and $\sim$ 60\%  of Q$^{3}$ and Q$^{4} $ mixture. In this sample, the lack of Q$^{1}$ group shows a high degree of connectivity of the SiO$_4$ tetrahedra. These analysis thus show that the amorphous samples have a large range of connectivity, on average larger than in the associated crystalline structure (Q$^{0}$ for forsterite and Q$^{2}$ for pyroxene enstatite and diopside). In addition, the F samples (Mg$_{2-2.8}$SiO$_4$) display a degree of connectivity lower than the E and D samples because the dominant species are Q$^{0}$ and Q$^{1}$  while it is Q$^{1}$ and Q$^{2}$ for the E sample, and Q$^{2}$ and Q$^{3}$  for the D sample.

\section{Infrared and submillimetre spectroscopy}
\label{spectro}

\subsection{Sample preparation}
\label{sample-prep}
To perform the measurements in the MIR spectral range, we mixed our silicate powder with KBr powder (Aldrich) in a mortar and we applied on this homogeneous mixture a pressure of 7~tons during 10~min, thus obtaining a 10~mm diameter pellet. Typically, we use 0.2-0.5 mg of samples embedded into 100~mg of KBr.\\
For the FIR spectroscopic measurements, the silicate grains were mixed with polyethylene (PE) powder at 120$ ^{\circ} $C and we applied on this homogeneous mixture a pressure of 7~tons during 5~min to obtained 10~mm diameter pellets.
The mass absorption efficiency (MAC), or opacity, of the dust analogues decreases with increasing wavelength. Therefore, in order to cover the spectral range from 100~$\mu$m to more than 1~mm, it is necessary to perform measurements on pellets containing more or less material depending on the spectral changes of the MAC. Typically, two or three pellets were needed, containing from 10~mg up to 100~mg of samples and 100~mg to 250~mg of PE.

\subsection{Experimental set-up}
\label{setup}
The MIR/FIR/submm optical properties of the samples were measured with the experimental set-up ESPOIRS (Etudes Spectroscopiques des Propri\'et\'es Optiques InfraRouge et Submillim\'etrique of interstellar dust analogues). It is composed of a Bruker IFS113V Fourier Transform InfraRed (FTIR) spectrometer coupled with a continuous flow helium cryostat which allows us to cool down the samples from 300~K to 10~K. The MIR transmission spectra were recorded using a KBr beamsplitter, a Globar source and a Deuterated TriGlycine Sulfate (DTGS) detector with a KBr window at a spectral resolution of the instrument of 1~cm$^{-1}$. In the MIR spectral range, the transmission measurements were performed only at ambient temperature for all the samples. \\To perform measurements in the FIR/submm spectral range, the spectrometer is equipped with a Hg lamp source, different Mylar beamsplitters with thickness of 23, 50, 125~$\mu$m and a 4~K Ge-bolometer detector. This configuration allows us to perform measurements in the 100 - 800~$\mu$m wavelength range (5000 - 12.5~cm$^{-1}$). To complete our wavelength range above 800~$\mu$m we performed experiments on the AILES (Advanced Infrared Line Exploited for Spectroscopy) beamline at the synchrotron SOLEIL \citep{brubach2010}. The experimental set-up is composed of a FTIR spectrometer Bruker IFS125 coupled with a a Helium pulse tube cryostat (Cryomech PT405). Thanks to the use of a 1.6~K bolometer and to the high brightness of the synchrotron IR light (100-1000 times greater than thermal source) we extended the spectral coverage to the lowest energy range, up to 1500~$\mu$m (6.6~cm$^{-1}$), with a very good signal/noise ratio. The sample holder in the two cryostats has two positions allowing us to record the transmission spectra of the sample and the blank PE pellet during the same cooling cycle. The transmittance spectra were recorded at 300, 200, 100, 30, and 10 K. 

\begin{figure*}[!t]
\includegraphics[scale=.37, angle=90]{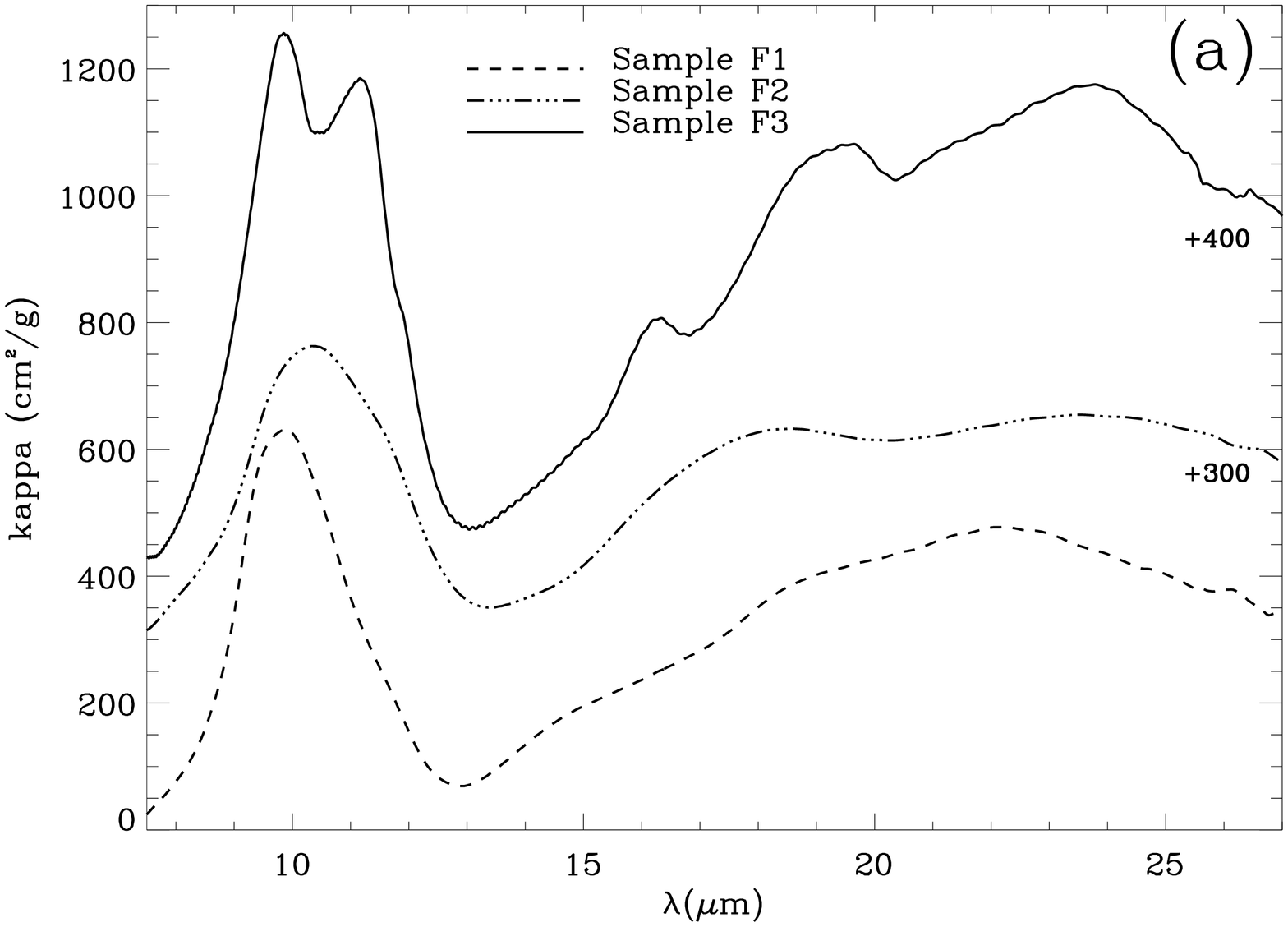}  
\includegraphics[scale=.37, angle=90]{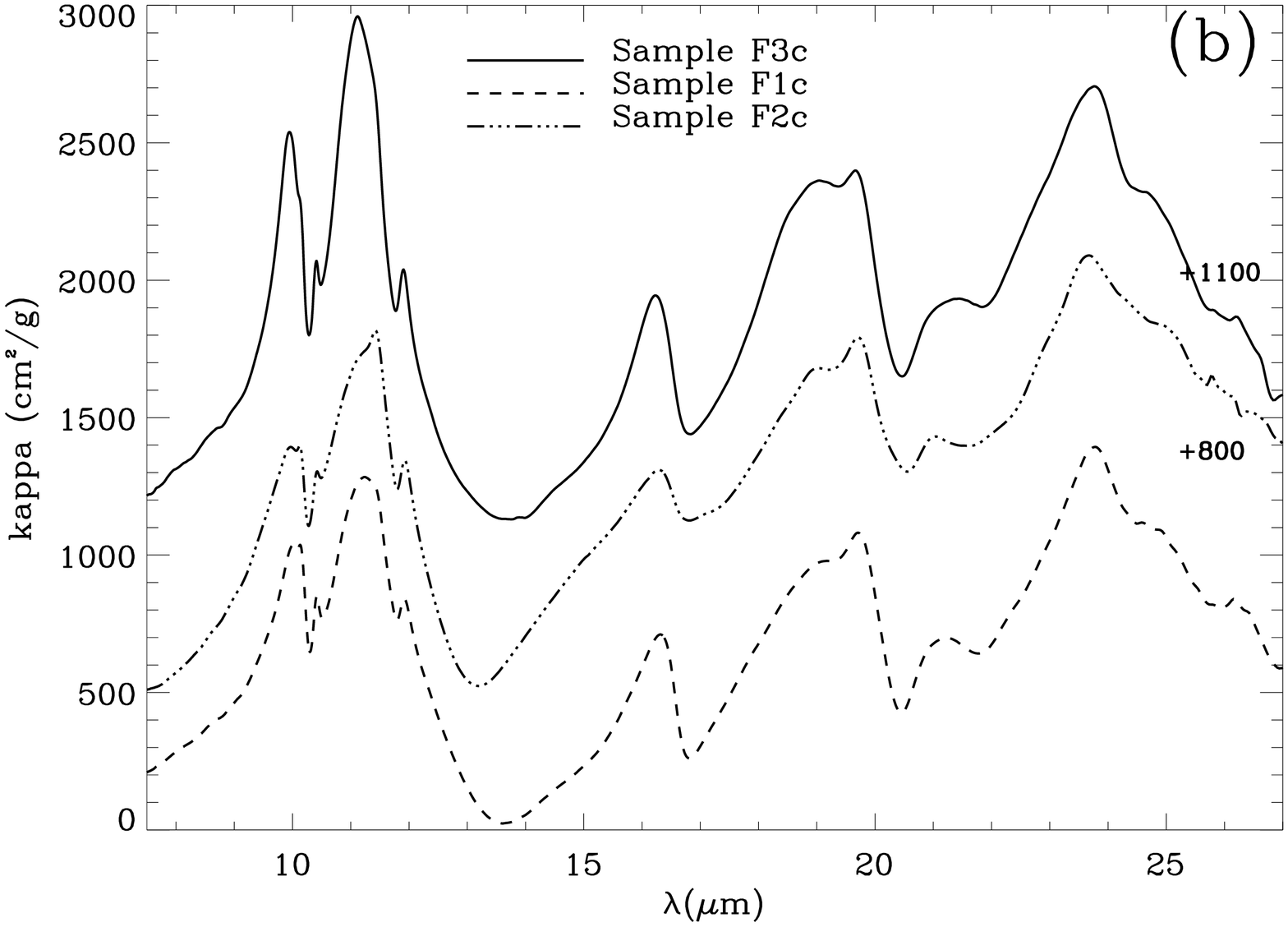}
\includegraphics[scale=.37, angle=90]{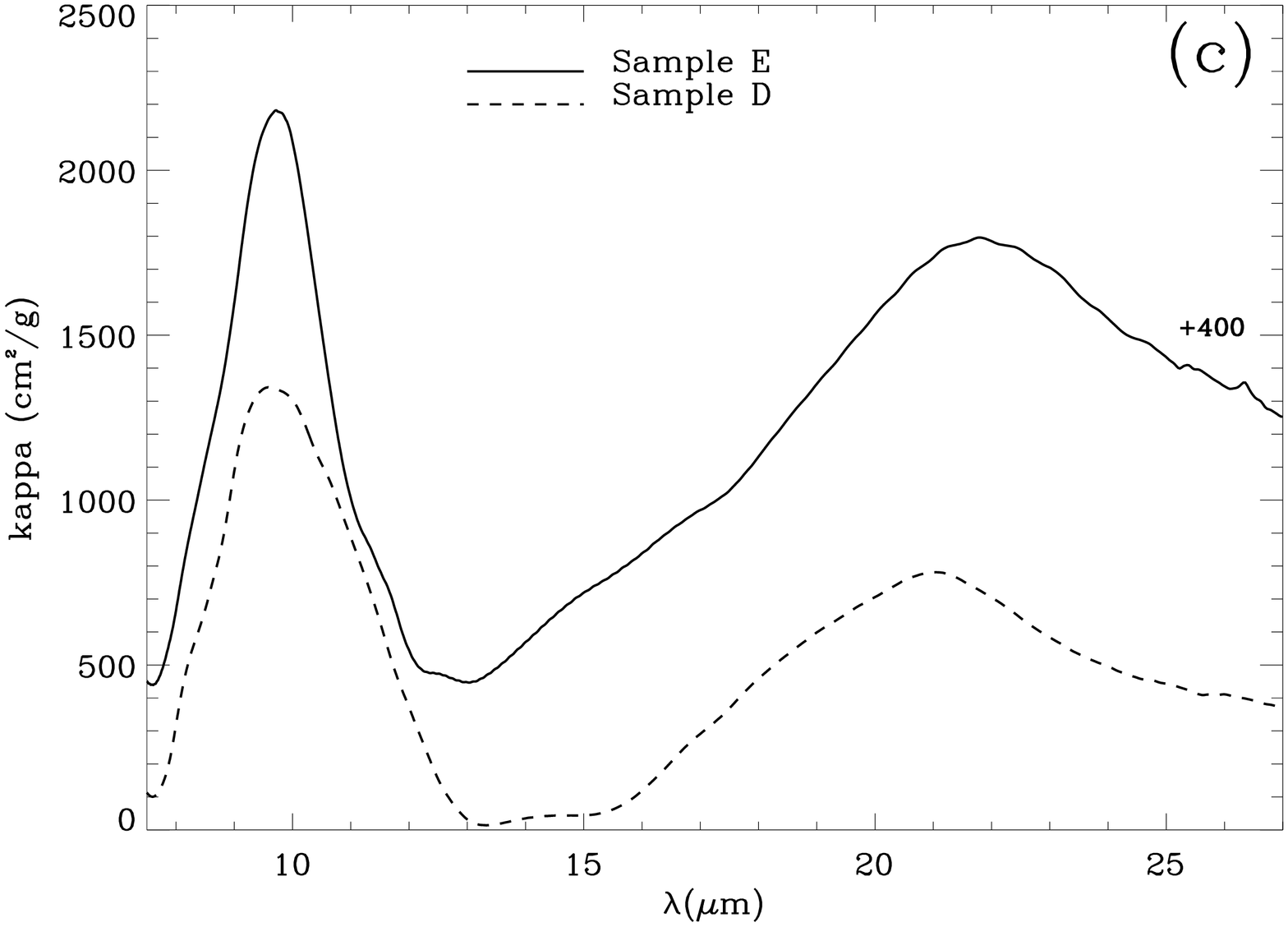}
\includegraphics[scale=.37, angle=90]{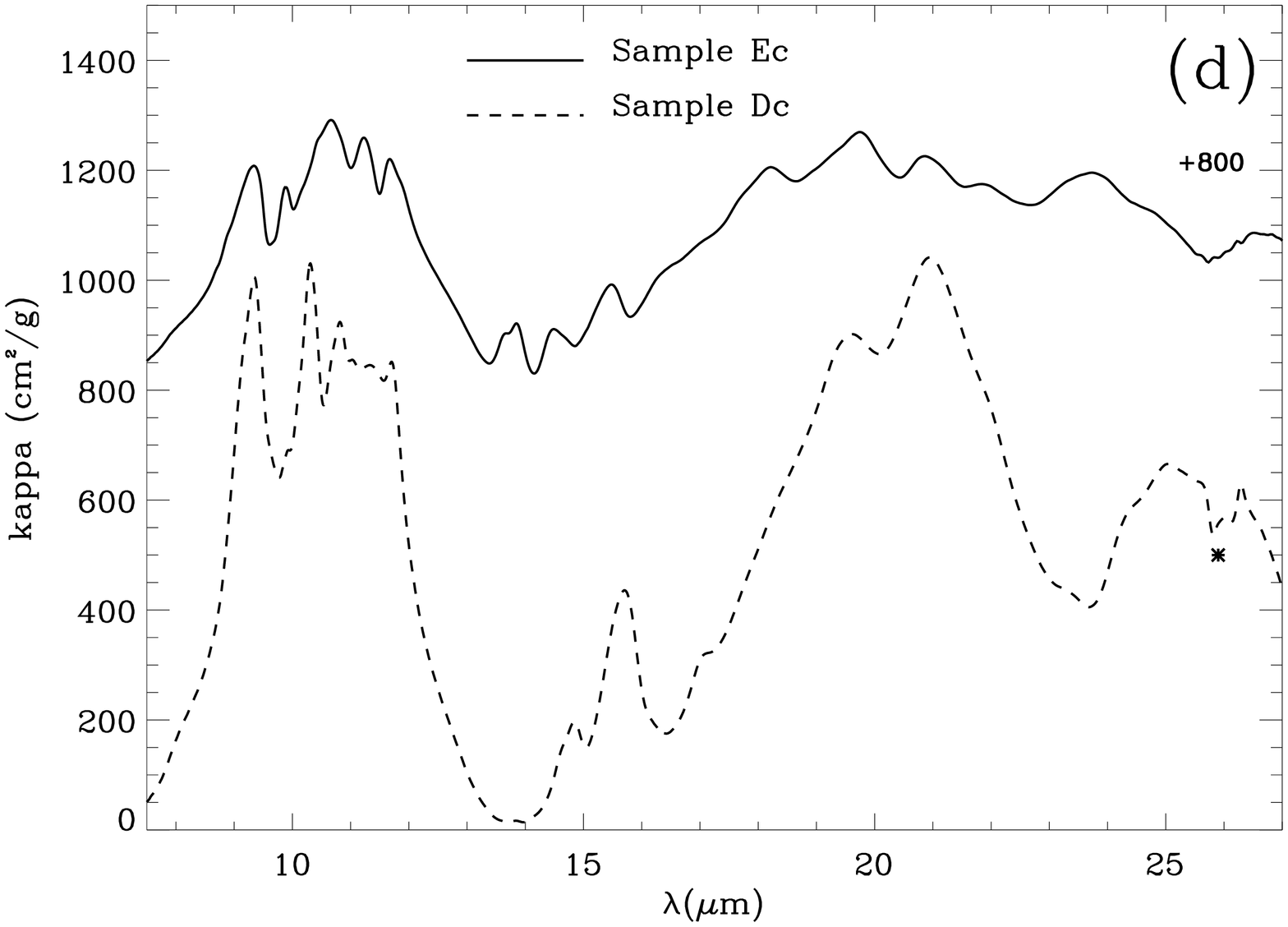}
       \caption{Mass absorption coefficient of the synthesised samples in the mid infrared range. Panel (a): amorphous samples with olivine stoichiometry, F1, F2, F3, panel (b) crystallised F1, F2 and F3 samples, panel (c) amorphous samples with enstatite and diopside stoichiometry, E and D, respectively, panel (d) crystallised E and D  samples. In panel(d) the absorption-like feature marked by an asterisk is an experimental artefact.}
    \label{mir}
\end{figure*}

\begin{figure*}[!t]
  \includegraphics[scale=.52, angle=0]{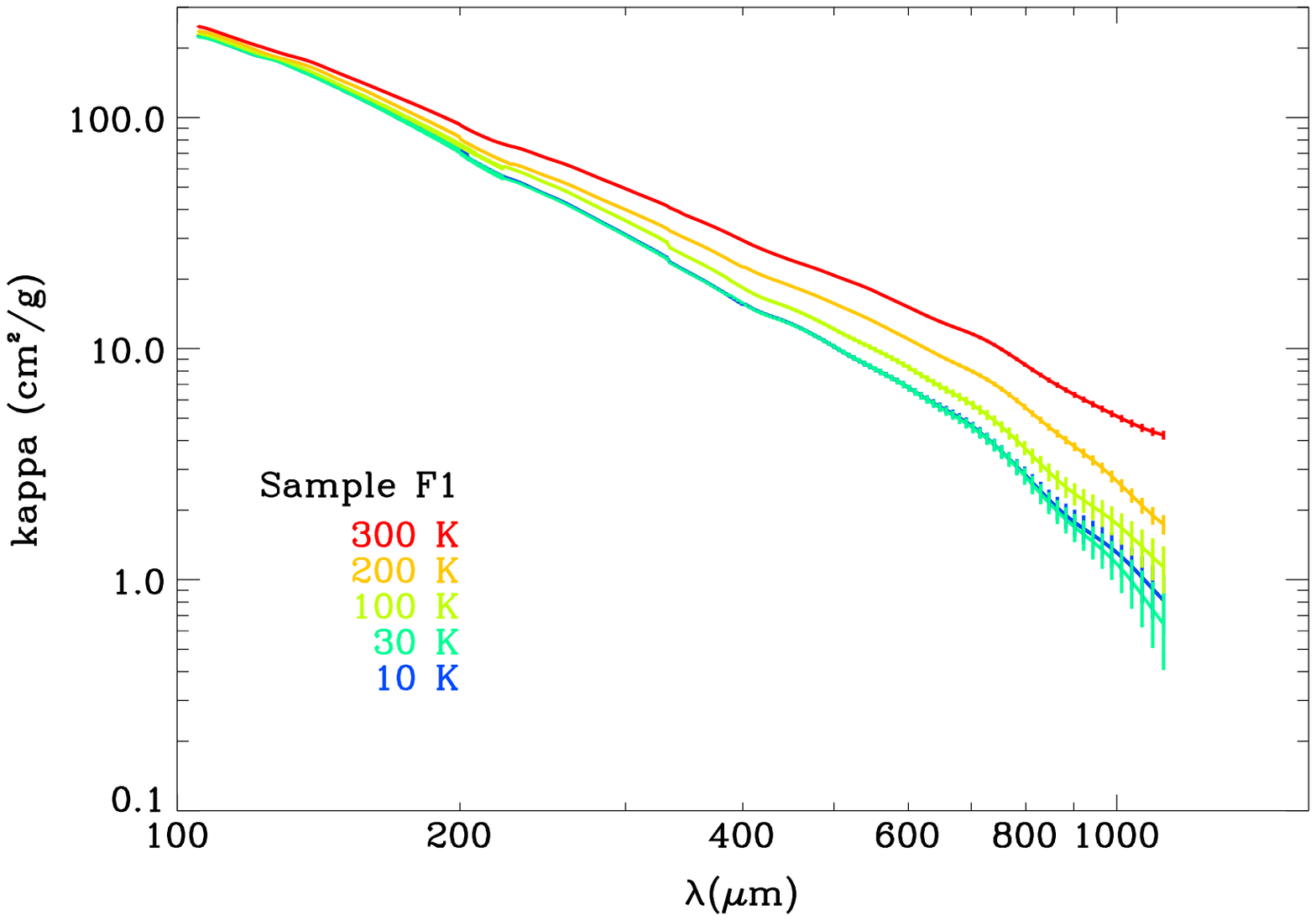}
  \includegraphics[scale=.52, angle=0]{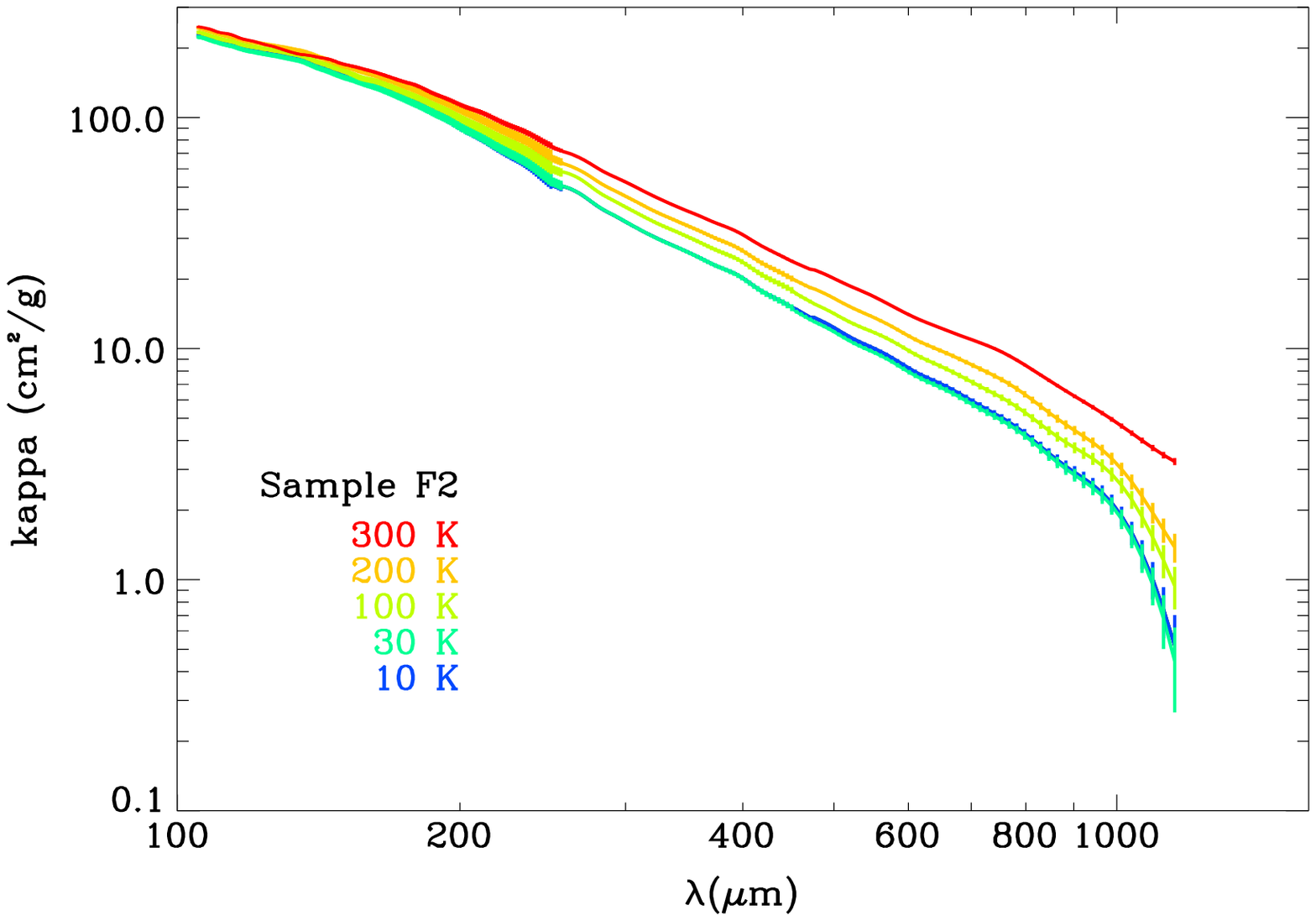}
  \includegraphics[scale=.52, angle=0]{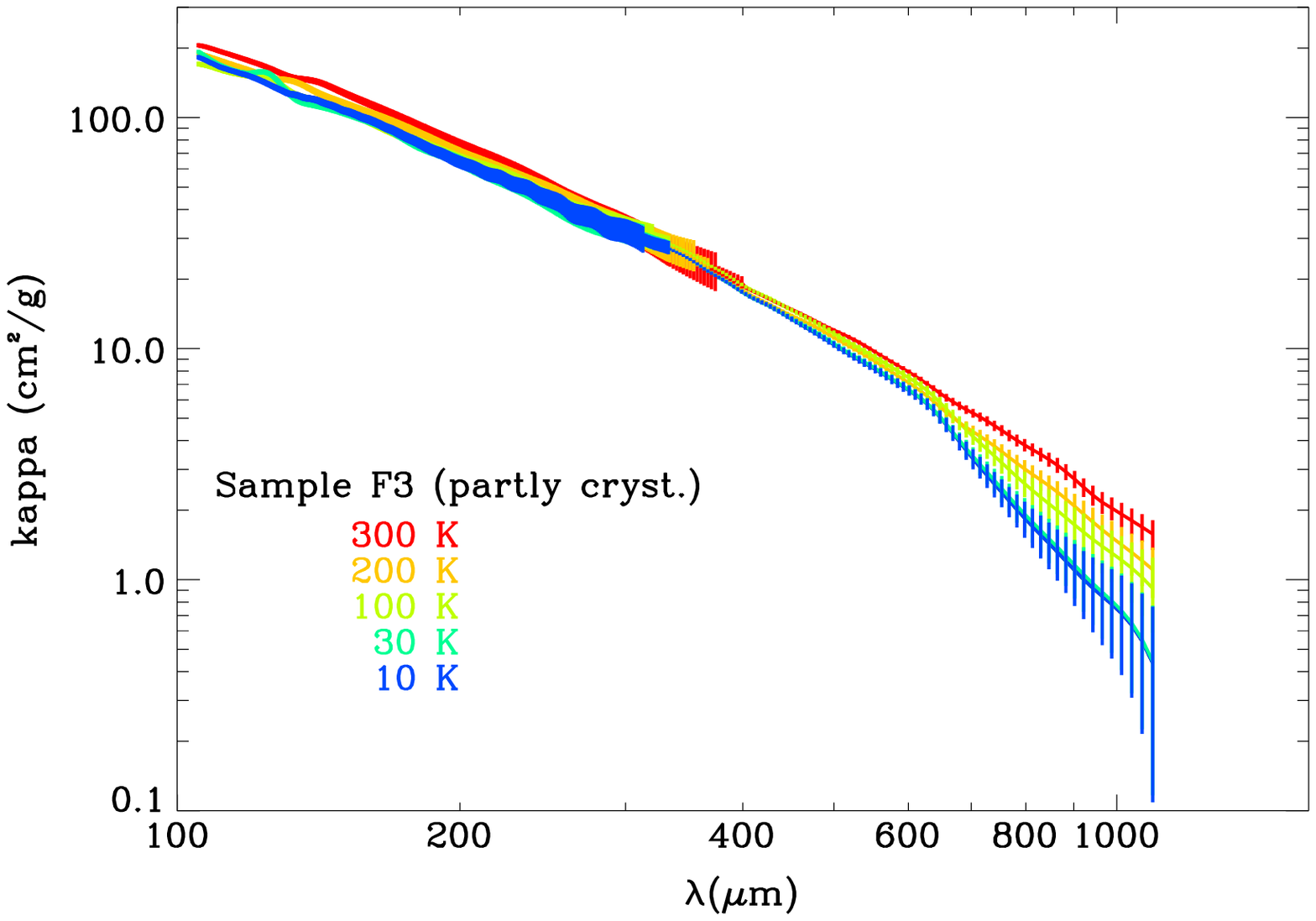}
    \caption{Mass absorption coefficient of amorphous olivine samples at different temperatures in the 100-1000(1200) $\mu$m range. Panel (a): sample F1: Mg$ _{2.3}$SiO$ _{4}$, panel (b): sample F2: Mg$ _{2.8}$SiO$ _{4}$, panel (c) sample F3: Mg$ _{2.05}$SiO$ _{4}$. }
    \label{F-FIR-col}
\end{figure*}

\subsection{Data analysis}
\label{data-analysis}
The transmittance spectra are calculated by dividing the spectrum of the sample pellet by the one of a blank pellet containing the same mass of matrix (KBr or PE depending on the spectral range) as the sample pellet. Hence, the blank and sample pellets do not have the same total mass nor the same thickness. The transmission of PE does not vary with the temperature from 300 K down to 10 K except around $\sim$ 125 $\mu$m where PE exhibits an absorption band which shape is temperature dependent. The transmittance spectra are affected by two problems. Firstly, they present very noisy regions that correspond to null absorption of the 23 $\mu$m beamsplitter around $\sim$ 155-180 $\mu$m. In this spectral region, and around $\sim$ 125 $\mu$m, we replaced the experimental data by a linear interpolation. Secondly, the multiple reflexions of the incident FIR beam between the parallel faces of the PE-pellets cause interference fringes that are most important at long wavelength (typically above 160 $\mu$m) where the material absorbs weakly. We applied two different methods to remove these fringes. The first method is similar to the one used by \citet{boudet2005} and consists in performing a polynomial fit of the transmittance spectra. We used the following function: $ f(x) = ax^6 + bx^5 + cx^4 + dx^3 + ex^2 +fx + g + h/x $. The second method is based on the convolution of the transmittance spectra with a sinc function (C. Tucker, private communication ).

The MAC of the pellet (sample and matrix), $\mathrm{{\kappa}_{pellet}}$ (cm$^2$.g$^{-1}$), is calculated from the transmittance spectra with the following formula (e.g. \citet{bohren1998}): 
\begin{equation}\label{kappa-t}
\mathrm{{\kappa}_{pellet} = - \frac{S}{M} \times ln(T)}
\end{equation}
where S is the pellet section (cm$^2$), M the sample mass in the pellet (g) and T the transmittance spectrum. We have assumed that scattering effects may be neglected since the grains size is much smaller than the wavelength. The MAC of the grains, $\kappa$, is related to the MAC of the pellet by the following relation (Eq. (6) from \cite{mennella1998}): 
\begin{equation}\label{kappa-g}
\mathrm{{\kappa}(\lambda, f) = g(\epsilon, f) * {\kappa}_{pellets}(\lambda, f)}
\end{equation}
where f is the volume filling factor of the grains within the matrix and g($\epsilon$, f) is the finite concentration reduction factor. The dilution factor adopted to make the pellets is high enough to ensure that the filling factor f never exceed 0.15 for all the pellets. Hence we can use Eq.(5) from \cite{mennella1998} to calculate g($\epsilon$, f). We have adopted the same value as \cite{mennella1998} for the dielectric function of the host medium (PE) and for the grain material. However, adopting \cite{boudet2005} values does not changes the value of g($\epsilon$, f) by more than 2-3\%. We find that g($\epsilon$, f) is usually greater than 0.9 except for pellets containing 50 and 100 mg of samples for which it is equal to 0.78. The transmittance corrected for reflexion loss, $\mathrm{T_{corr}}$, is calculated using the formula: $\mathrm{T_{corr} = T_{measured} /  (1-R)^2 }$ with R, the reflectance of the pellets, expressed as $\mathrm{R = (1-n_{eff})^2 / (1+n_{eff})^2}$. The effective refractive index of the pellet, $\mathrm{n_{eff}}$, is related to the average dielectric function of the pellet, $\mathrm{{\epsilon}_{eff}}$, by the formula $\mathrm{n_{eff} = \sqrt{Re({\epsilon}_{eff})}}$. The correction factor for reflexion loss, (1-R)$^2$, is larger than 0.9 for all our pellets.

Once these corrections have been applied to the different spectra corresponding to different amount of samples and different experimental configurations, they are merged into a single spectrum by taking advantage of the overlapping regions of the different spectra. The spectra were merged with respect to short wavelengths measurements. Typically the absolute opacity of the different spectra to be merged agree within less than 10\% at short wavelengths and within $\sim$ 10-30\% in the 250-300 $\mu$m range, i.e. between experimental configurations using the 23 and 50 $\mu$m Mylar beamsplitters. Thus we estimate that the uncertainty on the absolute value of the MAC, $\kappa$, is of the order of 10 to 30\%. However, it is important to note that, the relative uncertainty on the MAC of a sample measured at different temperatures within an experimental run is much lower, of the order of 5\% at most. This uncertainty is mainly due to the overall stability of the spectrometers which is critical at very long wavelength. The spectrometer stability was evaluated by reproducibility measurements and by comparing overlapping regions of spectra taken with different configurations and also of spectra taken with different spectrometers (the one from the ESPOIRS setup and the one from the AILES beam line at the SOLEIL synchrotron). The measurements at very long wavelengths were performed on the ESPOIRS experimental set-up up to 800 $\mu$m and were merged with measurements performed on the AILES beam line at the SOLEIL synchrotron in the range 500-1500 $\mu$m. Above 800 $\mu$m the measurements were performed solely with the AILES beam line setup. However, thanks to the high brightness and stability of the FIR synchrotron radiation and to the good performance of the spectrometer, the relative uncertainty on the transmission does not exceed 5 \%. This was tested from measurements performed during runs separated by several months.  It is comparable to the transmission accuracy stated by \citet{mennella1998}, 5-10 \%, and  \citet{boudet2005}, 2\%. The absolute uncertainty on $\kappa$ is expressed as follow: $\delta\kappa$ = S/M $\times \delta$T/T, where $\delta$T/T is the relative uncertainty on the measured transmittance, which is estimated to 2.5 \% and 5 \%  for spectra taken with the 23 $\mu$m and with the 50 or 125 $\mu$m beamsplitters, respectively. 

\section{Results}
\label{results}

\subsection{Mid Infrared spectra}
\label{s-mir}
We report in Fig.~\ref{mir}, panel a and c, the MAC spectra of all the amorphous silicate samples in the 7-27 $\mu$m range. We observe the two characteristic bands around 10~$\mu$m and 20~$ \mu$m which are attributed respectively to the Si-O stretching and O-Si-O bending vibrational modes. For samples F1, F2, E and D, the two bands are broad and featureless confirming the amorphous nature of these samples. However, unlike the MIR spectra of the F1 and F2 samples, the MIR spectrum of F3 clearly shows vibrational bands from crystalline material (at $\sim$ 11, 16, 19 $\mu$m) superimposed onto the broad and featureless10 and 20~$ \mu$m bands. Thus, the F3 sample is a mixture of amorphous silicate with forsterite whereas all the other silicates are completely amorphous. From the $^{29}$Si NMR spectrum of this sample, we estimate the amorphous crystalline fraction to be $\sim$ 14\% of the sample mass (see Sect~\ref{nmr}). 

The bands from forsterite dominate the spectra of the F1, F2 and F3 samples annealed at 1100~$^{\circ}$C (Fig.~\ref{mir}, panel b). The spectra of the three samples F1c, F2c and F3c are very similar and representative of the forsterite spectrum. The position and strength of the main spectral features of sample F1c, F2c and F3c are in agreement with the spectra of natural or synthesised forsterite \citep{fabian2001}. This is true even for the F1c sample which contains a significant amount of enstatite (see Table \ref{t_xrd}). The spectra of the E and D sample annealed at 1100~$^{\circ}$C, Ec and Dc (Fig.~\ref{mir}, panel d) are dominated by the characteristic vibrational bands of respectively enstatite and diopside \citep{koike2000}.

\subsection{Far Infrared and temperature dependent spectra}
\label{s-fir}
This section describes the main results obtained from the spectroscopic measurements performed in the FIR/submm range\footnote{The mass absorption coefficients of the studied samples are available upon request to Karine Demyk (demyk@cesr.fr).}. As exposed in Sect.~\ref{intro} and discussed in more detail in Sect.~\ref{astro}, simple semi-classical models allow us to express the grain MAC in the FIR/submm spectral range with the following formula (same as Eq.~\ref{kappa}): \\

\begin{equation}\label{kappa2}
\kappa =\:\kappa(\lambda_0)\: ( \frac{\lambda}{\lambda}_{0}) ^{-\beta}
\end{equation}

The parameter $\beta$, which is the slope of the MAC, $\kappa$, in logarithmic scale, in the FIR/submm range, is similar to the emissivity spectral index used in astronomical studies. These models predict a single value for $\beta$ in the FIR/submm domain that is independent of the temperature. Although $\beta$ is not a usual parameter to describe experimental data, we use it in the present study for comparison with the astronomical emissivity spectral index.  \\

\begin{figure*}[!t]
 \includegraphics[scale=.52, angle=0]{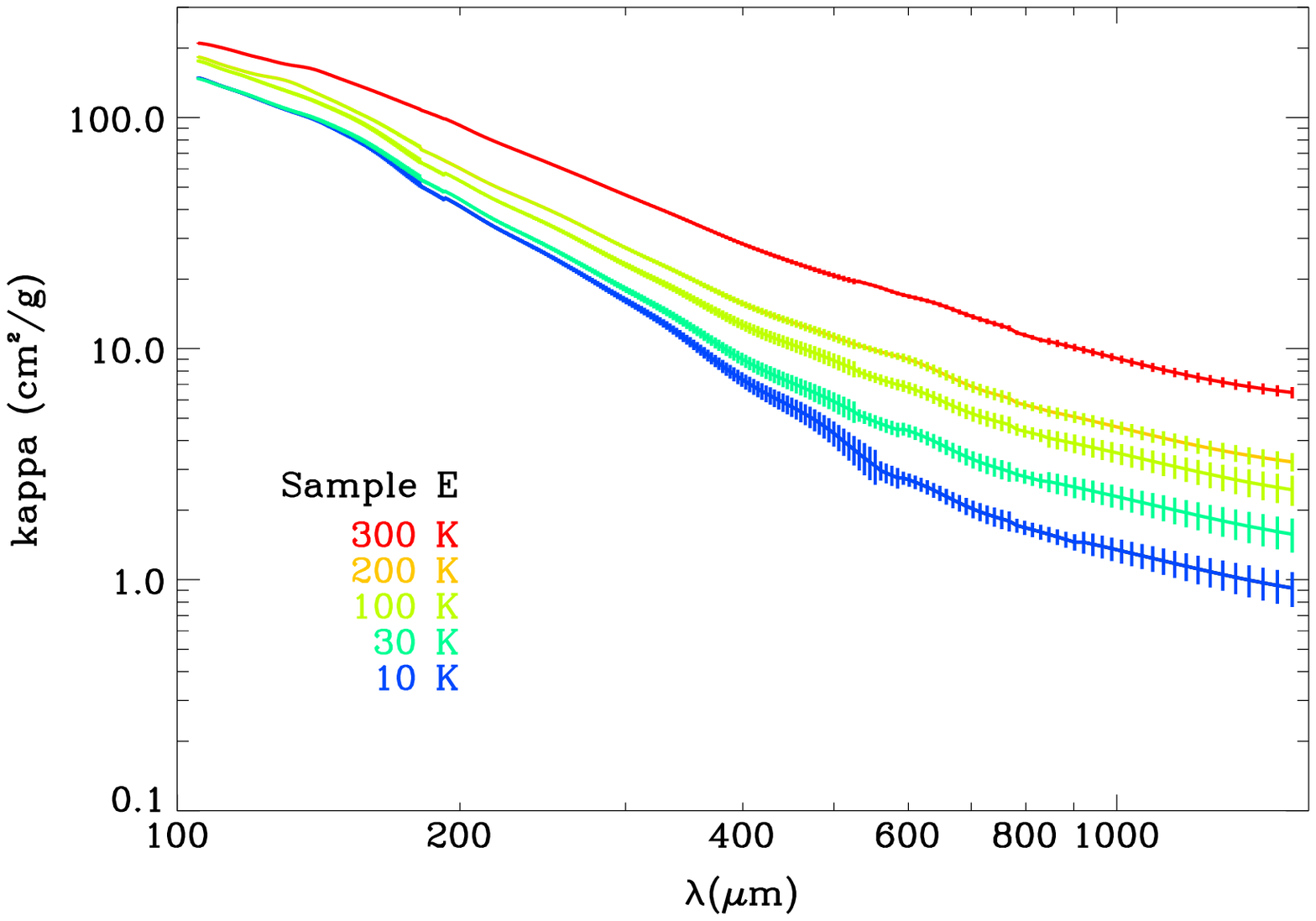}
    \includegraphics[scale=.52, angle=0]{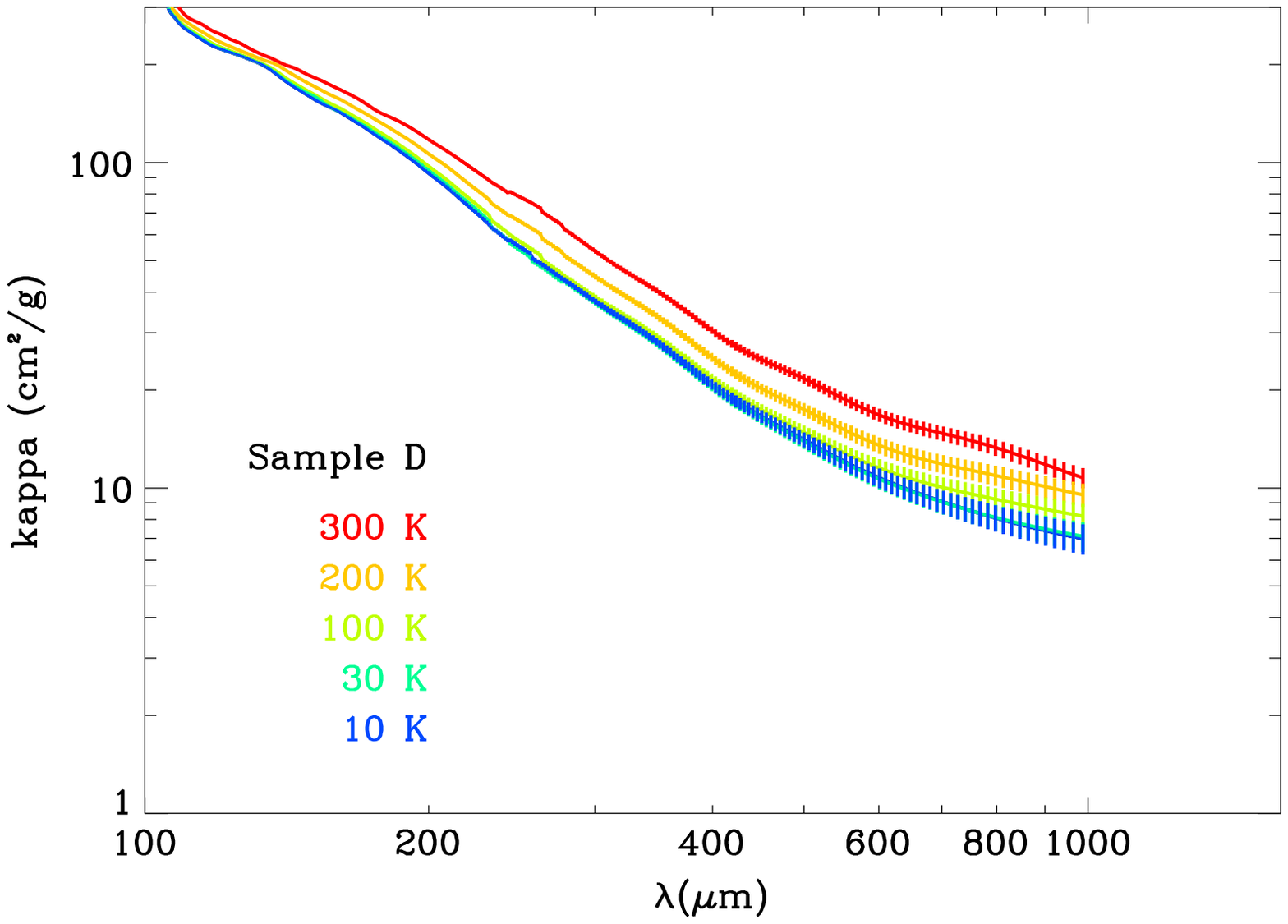}
      \caption{Mass absorption coefficient of amorphous samples with pyroxene stoichiometry at different temperatures in the 100-1000(1500) $\mu$m range. Panel (a): E sample, Mg$ _{0.95}$SiO$ _{3}$, panel (b): D sample,  Ca$ _{0.98}$Mg$ _{0.9}$Si$ _{2}$O$ _{6}$.}
    \label{ED-FIR-col}
\end{figure*}

\noindent {\textbf {\textit {Samples F1, F2 and F3 (Mg$_{\bf{2.05-2.8}}$SiO$_{\bf4}$):}}}\\
Fig.~\ref{F-FIR-col} shows the evolution of the MAC,  $\kappa$, of the amorphous samples F1, F2 and of the partially crystalline sample F3 as a function of $ \lambda $ in logarithmic scale and for different grain temperatures. Commonly for these three samples having an olivine stoichiometry, we observe that {\it (i)} the MAC diminishes when the grain temperature decreases, {\it (ii)} this diminution is more pronounced at long wavelengths and {\it (iii)} the shape of the MAC varies with the wavelength and its slope gets steeper at long wavelengths. For instance, the MAC at 1~mm, decreases by a factor of 2.1, 3.5 and 3.8 from 300 to 10~K, for the F2, F3 and F1 samples, respectively, whereas the decrease is negligible at 110 $\mu$m. We note that, for the 3 samples, the MAC does not vary for temperatures smaller than 30 K, the MAC measured at 10 and 30 K being almost indistinguishable. For all temperatures, the spectra show an inflexion in the absorption curves that is more or less pronounced, and that extend on a variable spectral domain depending on the considered sample. This break occurs around 550-700 $\mu$m depending on the sample but also depending on the grain temperature (it occurs at shorter wavelength at high temperature). It is therefore impossible to use Eq.~\ref{kappa2} to fit the MAC of these samples since a single spectral index cannot be defined over the entire FIR/submm range. However, on both sides of this inflexion point a linear behaviour is observed and we can thus define two spectral regions in which we can calculate $\beta$,  the slope of the curve (Table~\ref{beta-t}). For the F1 sample, at 300~K, the spectral index calculated in the 130-670~$ \mu$m range is $\beta$ = 1.6 whereas in the 710-1200 $ \mu$m range, $\beta$ =  2.2. At 10~K, $\beta$ is equal to 2.1 in the 130-690 $\mu$m range whereas it is 3.6 in the 690-1200 $\mu$m range. For the F2 sample at 300 K, in the range 170-770 $ \mu$m, $\beta$ =  1.8 whereas $\beta$ = 2.5 in the range 770-1000 $ \mu$m. At 10~K, in the same wavelength ranges, $\beta$ is respectively equal to 2.1 and 3.2. For the F3 sample at 300~K, the spectral index calculated in the range 140-550~$ \mu$m is equal to 1.9 whereas in the 550-1200 $\mu$m range it is equal to 2.5. At 10~K, $\beta$ is respectively equal to 1.9 and 4.5 in the wavelength ranges 150-650 $\mu$m and 650-1200$\mu$m. 

Even though all the three samples present a similar frequency and temperature dependence, some differences exist between their MAC. In the case of the F1 and F2 samples, we observe that the MAC decreases with decreasing temperature on the whole wavelength range whereas for the F3 sample, the variation of $ \kappa $ as a function of  temperature is present only for $\lambda \gtrsim$ 550-650 $\mu$m. This may be due to presence of crystalline material, forsterite, in the F3 sample. Interestingly, despite the crystallised fraction present in the F3 sample (14\% in mass, see Sects. \ref{nmr} and \ref{xrd}) and that is clearly observable in the MIR spectrum of this sample, its FIR optical behaviour is dominated by the 86\% amorphous fraction, as shown by the variation of its MAC with the temperature. The minor differences found for the values of $\beta$, for a given wavelength range and at a given temperature, between the remaining 2 samples F1 and F2 may be due to differences in composition (Mg excess), homogeneity and microstructure induced by the sol-gel methods used for the synthesis (see Sect.~\ref{sample}). \\

\begin{table*}[!ht]
\caption {Value of the  spectral index $ \beta $ derived on different spectral ranges from the experimental mass absorption coefficient spectra for the studied amorphous samples.} 
\label{beta-t}
\begin{center}
\begin{tabular}{llccccc}
\hline 
\hline 
Sample 				&  spectral domain ($\mu$m)\tablefootmark{(1)}	&  $\beta$ (10~K) & $\beta $ (30~K) & $\beta$ (100~K) &  $\beta$ (200~K) & $\beta$ (300~K) \\ 
\hline 
\hline 
F1 (Mg$_{2.3}$SiO$_{4}$)	& 130 - 690/710 		&  2.1	& 2.1	& 2.0	& 1.8	&	1.6  \\
                                                 & 690/710 - 1200 	&  3.6	& 3.8	& 3.4	& 3.2	&	2.2  \\
\hline
F2 (Mg$_{2.8}$SiO$_{4}$)	& 170 - 770/800 		&  2.1	& 2.2	& 2.1	& 2.0	&	1.8  \\
                                                 & 770/800 - 1000 	&  3.2	& 3.2	& 2.9	& 3.0	&	2.5  \\
\hline
F3 (Mg$_{2.05}$SiO$_{4}$)	& 150 - 550/650 		&  1.9	& 1.9	& 1.9	& 1.9	&	1.9  \\
                                                 & 550/650 - 1200 	&  4.5	& 4.5	& 3.4	& 3.0	&	2.5  \\
\hline
E ( Mg$ _{0.95}$SiO$ _{3}$)	& 150 - 420/590 		&  2.5	& 2.2	& 2.1	& 2.0	&	1.7  \\
                                                 & 420/550 - 800 		& 1.7	& 1.5	& 1.5	& 1.4	&	1.3  \\
                                                 & 800 - 1500 		& 0.9	& 0.9	& 0.9	& 0.9	&	0.9  \\
\hline
D (Ca$ _{0.98}$Mg$ _{0.9}$Si$ _{2}$O$ _{6}$)	& 150 - 450 			&  2.0	& 2.1	& 2.0	& 1.9	&	1.8  \\
                                                 & 450 - 650 			& 1.4	& 1.4	& 1.3	& 1.3	&	1.3  \\
                                                 & 650 - 1000 		& 0.8	& 0.7	& 0.6	& 0.6	&	0.9  \\
 \hline
\hline 
\end{tabular} 
\tablefoot{ \\
\tablefoottext{1}{The number of spectral domains on which a spectral index, $\beta$, may be defined depends on the samples. In addition, for a given sample, the spectral domains may vary with the grain temperature. When two values of the wavelength separating the spectral domains are given, the first one corresponds to the highest temperature (300 K) and the second one to the lowest temperature (10 K), see text.\\} 
}
\end{center}
\end{table*}

\noindent {\textbf {\textit {Sample E (Mg$_{\bf0.95}$SiO$_{\bf3}$):}}}\\
For the E sample, we observe that {\it (i)} the mass absorption coefficient decreases when the grain temperature decreases, {\it (ii)} the diminution of the MAC with the temperature is more pronounced at $\lambda \gtrsim$ 400~$ \mu$m, and {\it (iii)} the shape of the mass absorption coefficient varies with the wavelength and its slope gets flatter at long wavelengths (Fig.~\ref{ED-FIR-col}, panel a). Indeed, we note that at 200 $\mu$m, the mass absorption coefficient measured at 10~K is lowered by a factor of 2.2 with respect to the value at 300 K. This factor is equal to 3.9 at 400~$ \mu$m, to 6.8 at 800~$ \mu$m and remains constant up to 1500 $\mu$m. In contrast to the F samples, the MAC of the E sample varies significantly between 10 and 30 K. As for the F samples, the shape of the MAC clearly does not follow the asymptotic behaviour predicted by the semi-classical models: a single $\beta$ value cannot reproduce the experimental spectra. Unlike the F samples for which the MAC gets steeper, the MAC of the E sample is flattened at long wavelength. It is not possible for this sample, to define only two regions characterised by a linear behaviour with the wavelength. Even though it is more difficult to define separated spectral regions than for the F samples, we identify three regions in the spectra for which we can define and calculate a spectral index. Each region is delineated by an inflection in the absorption curves at 420-590 $\mu$m and at 790-800~$ \mu$m, depending on the grain temperature. The spectral index, $\beta$, calculated in these 3 wavelength ranges is indicated in Table~\ref{beta-t}. We note that at a given grain temperature, $ \beta $ decreases when $ \lambda $ increases. The spectral index derived from the MAC measured at 300~K is $\beta$ = 1.7 in the 150-420$\mu$m range, it decreases to 1.3 in the 420-800 $\mu$m range and to 0.9 in the 800-1500 $\mu$m range. We observe the same flattening for the spectral index derived from the MAC measured at 10~K: $\beta$ = 2.5, 1.7 and 0.9 in the wavelength ranges: 150-590, 590-790 and 790-1500 $\mu$m, respectively. At a given wavelength, $\beta$ increases when the temperature decreases only for the wavelength range $\lesssim$ 800~$ \mu$m. Above 800~$ \mu$m, $ \beta $ is constant with the temperature.\\ The MAC of the E sample is in qualitative agreement with the MgSiO$_3$ sample studied by \citet{boudet2005} even though some differences exist that may be attributed to differences in the samples arising from different synthesis methods.  In both experiments, the MAC is characterised by a flattening at long wavelength. However, the comparison between the two measurements is not straightforward. Indeed, the values \citet{boudet2005} derived for $\beta$ are not comparable with ours because they are calculated in two different spectral ranges (100-200 $\mu$m and 500-1000$\mu$m) that are not adapted to the sample E presented in this study. This illustrates the problem of the relevance of the use of the spectral index $\beta$ (see Sect.\ref{beta-model}).\\

\begin{figure*}[!ht]
   \includegraphics[scale=.52, angle=0]{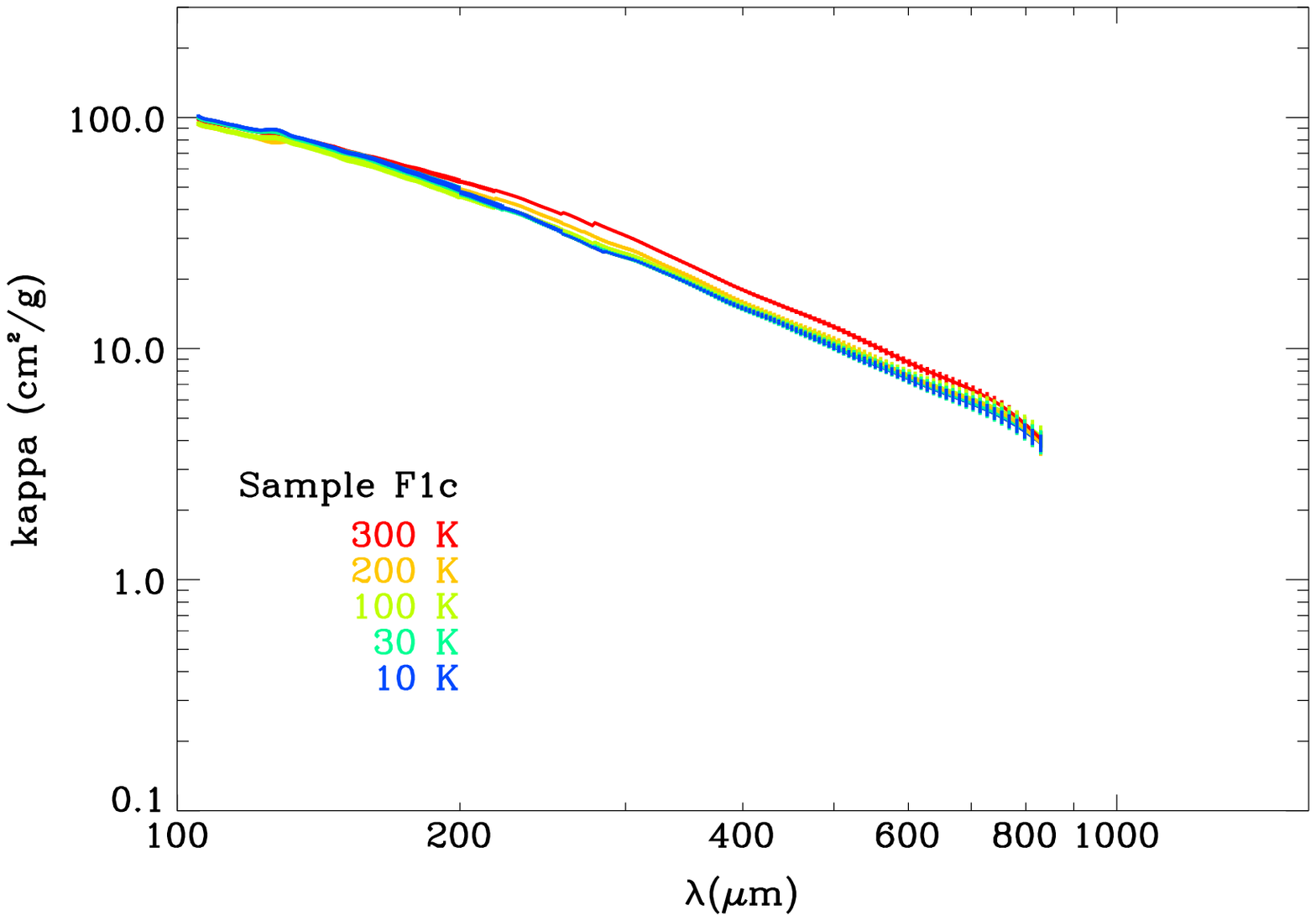}
   \includegraphics[scale=.52, angle=0]{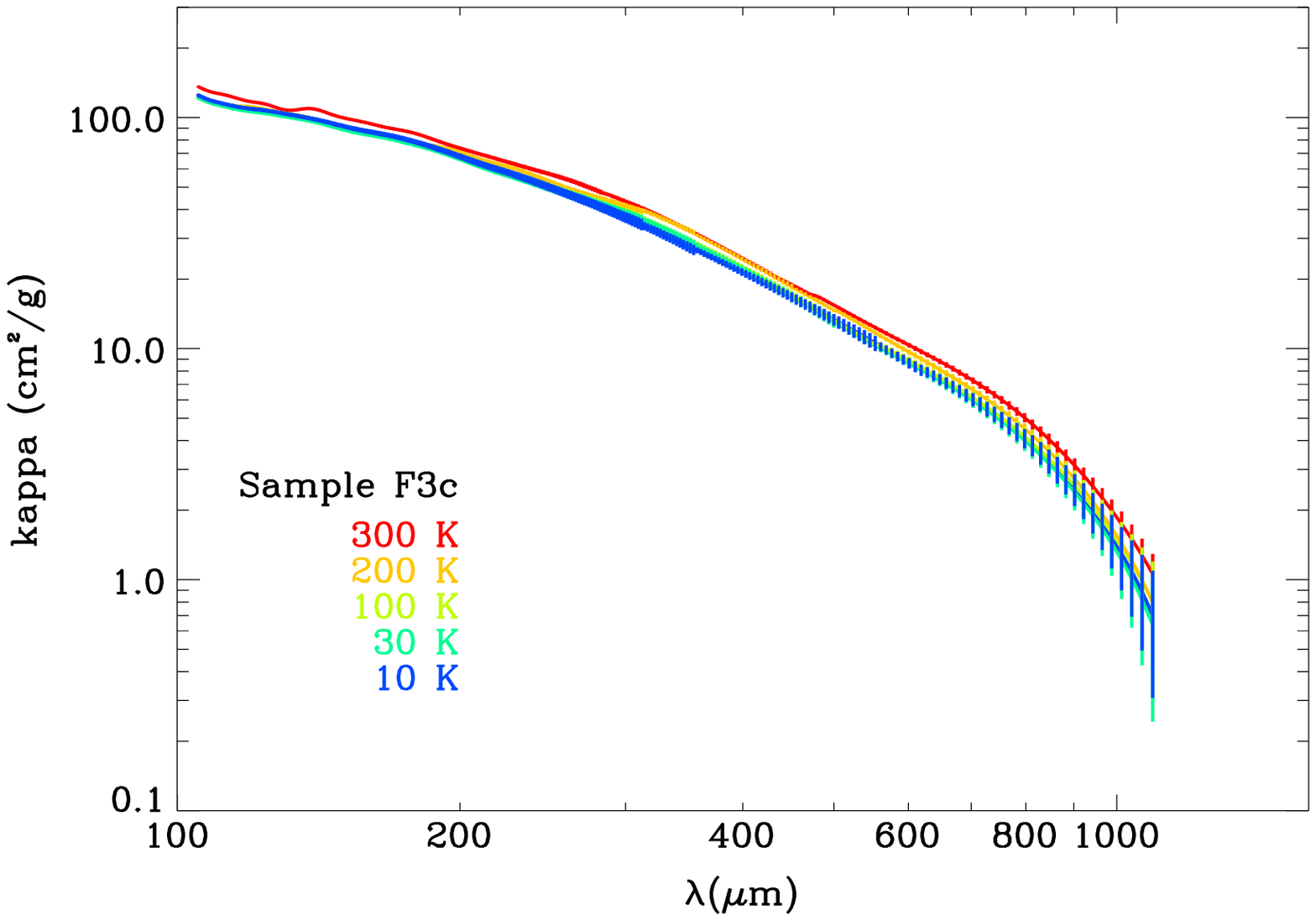}
    \includegraphics[scale=.52, angle=0]{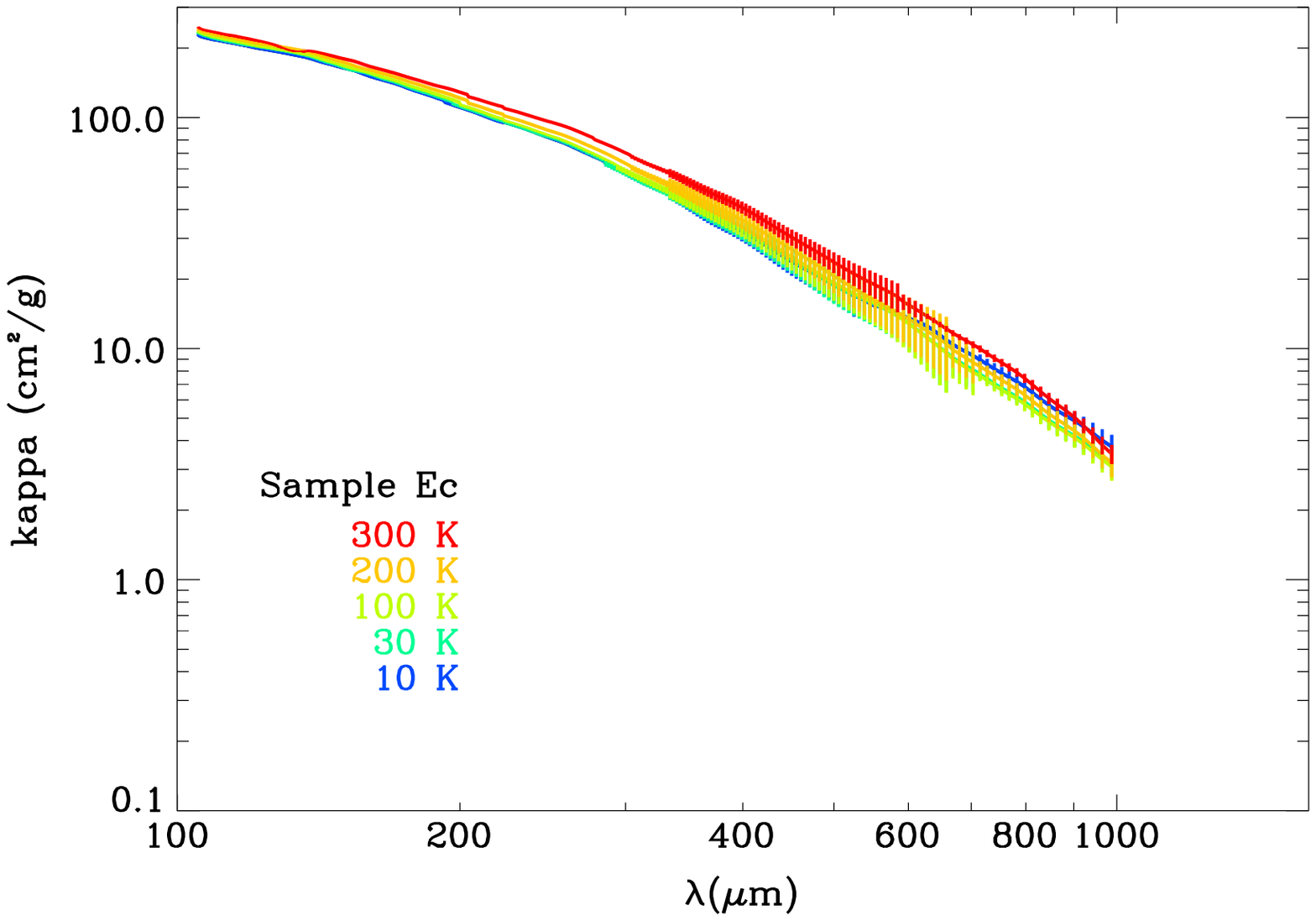}
    \includegraphics[scale=.52, angle=0]{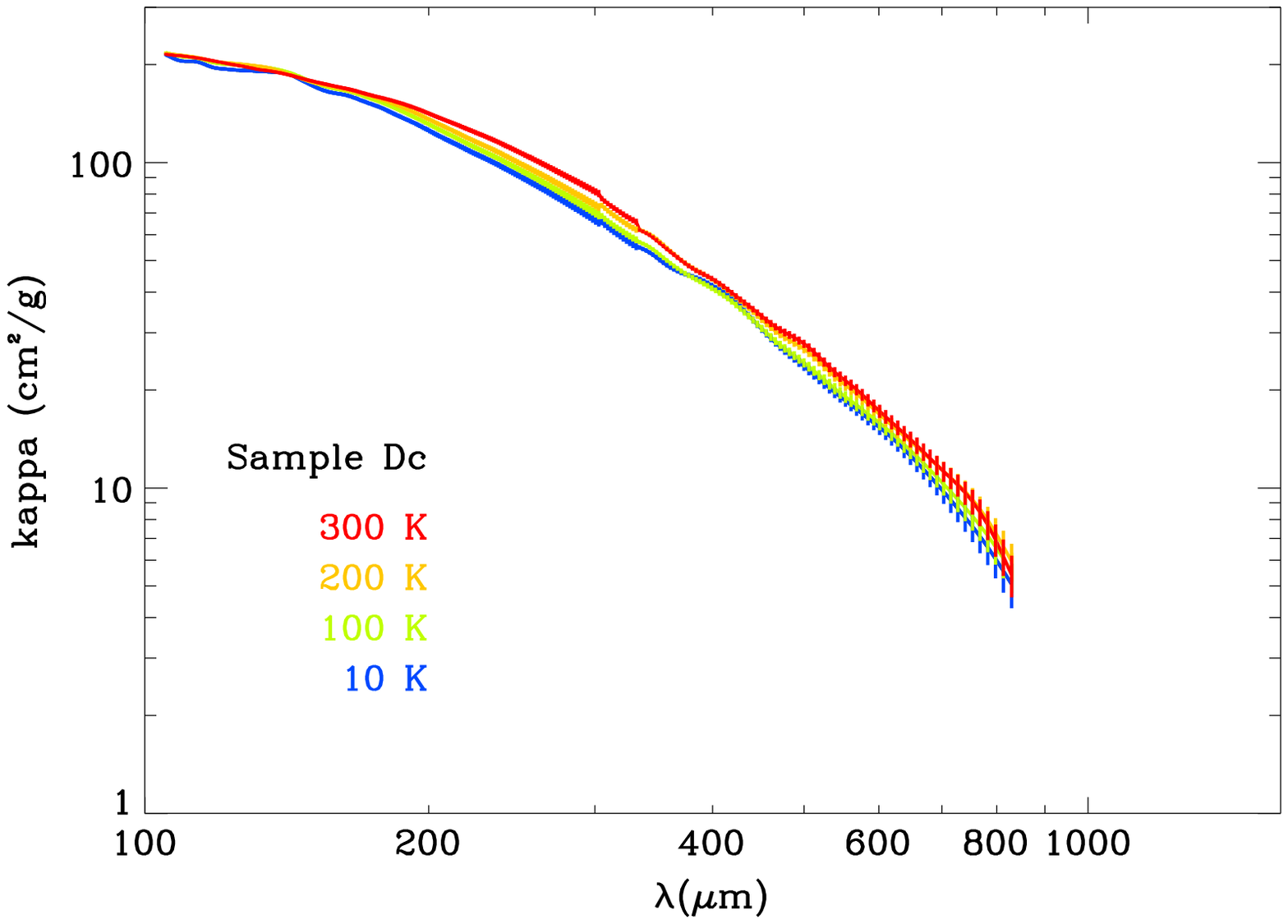}
      \caption{Mass absorption coefficient of crystallised olivine and pyroxene samples at different temperatures in the 100-1000(1200) $\mu$m range. Panels (a-b): Mg$_2$SiO$_4$ samples, panel (c) MgSiO$_3$ and panel (d) CaMgSi$_2$O$_6$.}
    \label{FIR-cris-col}
\end{figure*}

\noindent {\textbf {\textit {Sample D (Ca$_{\bf0.98}$Mg$_{\bf0.9}$Si$_{\bf2}$O$_{\bf6}$): }}}\\  
The variations of the MAC of the D sample with the wavelength and with the grain temperature (Fig.~\ref{ED-FIR-col}, panel b) is qualitatively similar to what is observed for the E sample: {\it (i)} decrease of the mass absorption coefficient with grain temperature that {\it (ii)} is more pronounced at long wavelength, and {\it (iii)} flattening of the MAC at long wavelength. However, quantitatively, these variations are much less pronounced in the case of the D sample. For the amorphous D sample, we observe the decrease of $ \kappa $ with decreasing temperature but this diminution is almost constant above $\sim$ 250~$ \mu$m. As for the other F and E samples, the MAC cannot be described by a single spectral index $\beta$. For all temperatures, the spectra are characterised by an inflection point in the absorption curves at $\sim$ 450~$ \mu$m and another at $\sim$ 650~$ \mu$m. Like for the E sample, these inflection points correspond to successive flattening of the MAC. Indeed, between 150 and 450~$ \mu$m, we calculate at 300~K a spectral index with a value of 1.8 that decreases to 1.3 between 400 and 600~$ \mu$m and then to 0.9 in the 650-1000~$\mu$m. Finally, we observe at a given wavelength, that $ \beta $ is constant with the temperature (Table~\ref{beta-t}). \\

\noindent {\textbf {\textit {Crystallised F1c, F2c, F3c, Ec and Dc samples:}}}\\
To investigate the hypothesis that the amorphous state could be responsible for the variation of the grain MAC as a function of  temperature and wavelength, part of each sample were heated for 24 hours at 1100$^{\circ}$C to crystallise them. At this temperature we know from the MIR (Fig ~\ref{mir}) and the NMR and XRD analysis that our samples are crystalline. We report the evolution of $ \kappa $ as a function of $ \lambda $ in logarithmic scale at different temperatures for the crystallised samples F1c and F3c (olivine stoichiometry) and for the crystallised samples Ec and Dc (pyroxene stoichiometry) (Fig.~\ref{FIR-cris-col}). The shape of the mass absorption coefficient deviates from the asymptotic form characterised by a single value for $\beta$. This may be due to the mixture of the different crystalline phases that are present in the annealed samples. During the process of crystallisation, separate phases have been formed. The F samples have crystallised in 2 or 3 phases depending on the sample (see Sect~\ref{xrd} and Table~\ref{t_xrd}): forsterite (Mg$_2$SiO$_4$), periclase (MgO) and ortho-enstatite (MgSiO$_3$), the E sample in 3 phases: ortho-enstatite, forsterite, and clino-enstatite and the D sample in 3 phases: diopside, akermatite and merwinite (see Table~\ref{t_xrd}). However it is important to note that, in contrast to the amorphous samples, the MAC of the crystallised samples does not vary with the temperature. The MAC value stays constant for all grain temperatures.  
In addition its spectral shape remains unchanged for all grain temperatures, i.e. the spectral index does not vary with the grain temperature. The optical properties of amorphous and crystallised samples thus behave very differently in the FIR/submm range as a function of the temperature.

\section{Discussion}
\label{discussion}
 
The study of the dust analogue MAC in the FIR/submm and of its temperature dependence reveals differences between the various materials (samples having an olivine stoichiometry: Mg$_{{2.05-2.8}}$SiO$_{4}$ versus samples with a pyroxene-stoichiometry: Mg$_{0.95}$SiO$_{3}$ and Ca$_{0.98}$Mg$_{0.9}$Si$_{2}$O$_{6}$), and at the same time a similar behaviour. The common behaviour can be characterised as follow:
\newcounter{qcounter}
\begin{list}{\arabic{qcounter}.~}{\usecounter{qcounter}}
\item All amorphous samples present a MAC that increases when the temperature rises, and this variation vanishes when the samples are crystallised by annealing. This implies that thermally-activated processes causing an additional absorption exist only in the amorphous state. When the temperature increases, the thermally activated processes increase the MAC, first in the long wavelength range ($\lambda\gtrsim$ 500-600 $\mu$m) at low temperatures and then progressively over the whole FIR/mm range at higher temperatures.
\item At the lowest experimental temperatures, when these thermally-activated processes do not add a significant absorption, the remaining temperature-independent MAC presents a common behaviour in all our amorphous samples: in the short wavelength range (100 $\lesssim \lambda \lesssim$ 500-700 $\mu$m), the local spectral index, has a value of the order of 2, whereas at longer wavelength, the value of $\beta$ varies after a more or less abrupt wavelength transition (Table~\ref{beta-t}).
\end{list}
The differences in behaviour of the MAC of the amorphous samples of different chemical compositions (olivine and pyroxene composition) with the wavelength and temperature are the following:

\begin{list}{\arabic{qcounter}.~}{\usecounter{qcounter}}\addtocounter{qcounter}{2}
\item The F1, F2, and F3 samples (olivine stoichiometry) present a MAC characterised by a spectral index around 2, temperature-independent in the 120-650 $\mu$m range, followed by a steep transition towards spectral index values greater than 3 at $\lambda \ge$ 700 $\mu$m. Despite the differences in composition due to different synthesis recipes and despite the presence of a small amount of crystalline material in one of the sample, the behaviour of the 3 samples with an olivine composition samples are similar. 
\item The behaviour of the 2 samples with a pyroxene composition, E and D, is different from the one of the samples with an olivine composition. The spectral index value around 2 in the short wavelength range evolves with a smooth transition towards lower values, down to 1 for $\lambda \ge$ 800 $\mu$m. However, despite this similar behaviour, the variation of the MAC with the temperature is much higher in the E sample than in the D sample.\\
\end{list}

\noindent From point (1) it is clear that the variation of the MAC with the grain temperature is related to the disorder of the material, i.e. to its amorphous nature. This is clearly shown by the disappearance of these variations when the materials are crystallised, i.e. well-ordered. This is true for all samples: the F1, F2, and F3 samples with an olivine stoichiometry, the E and D samples with a pyroxene stoichiometry as well as for silica samples (Coupeaud et al. in preparation). A similar behaviour was also observed by \citet{mennella1998} for iron-rich amorphous and crystalline samples with an olivine stoichiometry  (Mg$_{1.12}$Fe$_{1.88}$SiO$_4$). Hence, the MAC variation with the grain temperature is a general characteristic of amorphous materials, regardless of their composition or detailed micro-structure. However, the amplitude and wavelength dependence of this variation is different for the various samples (olivine- versus pyroxene- stoichiometric samples but also within the pyroxene- or olivine stoichiometric samples). The origins of these differences are likely to be linked with the structural state of the amorphous samples which differs by their compositions and preparation methods. The detailed description of an amorphous phase is difficult as well as to identify measurable physical parameters that may be responsible for the observed MAC variations. Crystals are solely described by their composition and lattice periodicity but this is not the case for amorphous solids that are characterised by disorder of chemical nature (at a given position into the solid the composition of the atom changes) and of geometrical nature (the bond lengths and angles vary). Even though they may have broad structural order at the short (nearest neighbours) and intermediate (second or higher coordination sphere) length scales, amorphous solids are characterised by the lack of long-range periodicity. Consequently, in amorphous solids there is flexibility in the way that atoms can be organised in a variety of local arrangements, thus resulting in a diversity of structures at short and medium length scales. Composition is certainly a parameter that may be used to describe amorphous solids since it has a strong influence on the atomic arrangement at short- and intermediate-range order but it is not the only one since amorphous materials having the same composition may have different microscopic structures as this is the case for glass and solgel materials.

\noindent If one considers only the spectral shape of the MAC of the studied samples at a given grain temperature, the experiments presented in Sect.~\ref{results} show that the FIR/submm MAC of the F1, F2 and F3 samples and of the E and D samples are different (points 3 and 4). Interestingly, experiments performed on silica samples, SiO$_2$, have revealed a spectral shape of the MAC of silica different from the olivine and pyroxene samples (Coupeaud et al, in preparation). Hence, this seems to indicate that the spectral shape of the MAC of amorphous silicate samples depends, at first order, on the cation/Si ratio of the material, which is equal to 0 for the silica samples, (Mg$^{2+}$+Ca$^{2+}$)/Si = 0.95 and 0.94 for the samples with a pyroxene stoichiometry, E and D samples, respectively,  and Mg$^{2+}$/Si = 2.05, 2.3 and 2,8 for the F3, F1 and F2, {samples with an olivine stoichiometry}, respectively. The cation/Si ratio seems to be a better parameter than the composition, since the samples E and D, have different composition but their MAC has qualitatively the same spectral shape.
This is also true for the 3 Mg-rich samples which MAC has also qualitatively the same spectral shape as the one of Fe-rich olivine-stoichiometry FAYA sample (Mg$_{0.18}$Fe$_{1.82}$SiO$_4$) studied by \cite{mennella1998}. However, because iron is an important component of interstellar dust it would be interesting to study more Fe-rich samples (with iron within the silicate network for different stoichiometries from enstatite to olivine, with oxide inclusions) to investigate whether its presence within the grains may induce or not a different spectral behaviour of the MAC.

\noindent However, for a given value of the cation/Si ratio, the amplitude of the MAC variations changes from one sample to another as this is the case for the E and D samples. These 2 samples have different compositions, with the D sample containing Ca$^{2+}$ cations in addition to Mg$^{2+}$ cations. The presence of Ca$^{2+}$ cations, which are linked to 6 oxygen atoms whereas Mg$^{2+}$ cations are linked to 4-5 oxygen atoms, modifies the chemical and geometrical microstructure of the D sample compared with the E sample and it certainly participates to explain the MAC variations. As far as the samples with an olivine stoichiometry are concerned, the differences between the 3 samples may also be explained by differences in their stoechiometry (the Mg$^{2+}$/Si ratio varies from 2.0 to 2.8) and thus in their chemical microscopic structures resulting from the use of different recipes for the solgel synthesis. Indeed the XRD analysis show that the crystallised samples F1c, F2c and F3c are a mixture of different crystalline phases and it is likely that the amorphous samples contain precursors of these phases such as for example chains of SiO$_4$ tetrahedra for the F1 sample (precursor of enstatite), more or less MgO inclusions in samples F1 and F2 and partly crystallised material in the F3 sample. 

\noindent The observed variations of the MAC may also be understood in term of degree of connectivity of the SiO$_4$ tetrahedra in the material (obtained from the NMR analysis). The samples having an olivine stoichiometry contain a majority of isolated tetrahedra (Q$^0$ species) and dimers of SiO$_4$ (Q$^1$ species) whereas the samples having a pyroxene stoichiometry contain longer chains (Q$^2$ species) and 3-dimension structures (Q$^3$ and Q$^4$ species). The sample having a stoichiometry of diopside, for which the MAC variation is smaller than for the sample having an enstatite stoichiometry, contains much more Q$^3$ and Q$^4$ species than the sample having an enstatite stoichiometry. The link between the connectivity of the material and the optical behaviour at low temperature should be investigated further via more sensitive and detailed NMR studies. 

\noindent Our samples, because of the synthesis method, contain OH groups and water that may influence the MAC variation of the silicates samples in a similar way as for silica samples \citep{agladze1996,boudet2005}. \citet{boudet2005} proposed that OH groups could be involved into tunnelling processes in two-level systems (TLS) explaining partly the experimental MAC variations in SiO$_2$ samples. From the OH vibrational bands at $\sim$ 3 $\mu$m, we estimate that all samples have roughly the same amount of water and OH groups. Hence, even though water and OH groups may participate to the temperature dependance of the MAC, they are certainly not responsible for the observed spectral shape of the MAC since the wavelength dependence of the MAC of the F samples, with an olivine stoichiometry, is different from that of the  E and D samples, with a pyroxene stoichiometry. As said above it is probably related to the cations (Mg$^{2+}$ and Ca$^{2+}$) and to their abundance with respect to silicon. 


\noindent It is interesting to note that the different spectral behaviours in the FIR range of the F samples compare to the E and D samples are accompanied by different absorption properties in the 8-12 $\mu$m range, i.e. in the Si-O stretching mode of the samples. The stretching modes of the E and D samples are very similar and peaks toward shorter wavelengths compared with the F sample stretching mode (Fig.~\ref{mir}). Hence, the MIR spectrum could be used to predict the FIR behaviour of the dust analogues. Unfortunately this may not hold for the astronomical silicates for which the MIR spectrum most certainly reflect a mixture of amorphous silicates with various compositions.

\noindent The common behaviour of all samples (points 1 and 2) comes from the general properties characteristic of the amorphous state. It may be understood in the framework of the TLS model developed within our group  \citep{meny2007}. The TLS model decompose the MAC in two components, one that depends on the grain temperature and the other that is temperature-independent. It predicts that the temperature-independent MAC in the FIR/mm spectral range is characterised by a local spectral index varying from $\beta$ = 2 at short wavelengths to $\beta$ = 4 at long wavelengths. The transition between these two values usually occurs in the 500-1000 $\mu$m range, depending on a characteristic length that describes the charge disorder in the amorphous material. This temperature-independent behaviour is expected to be observed at low temperature, typically, T $\sim$ 10 K. When the temperature increases, a thermally activated additional opacity appears, characterised by a lower value of the local spectral index that may be as low as 1. The amplitude of the variation of this additional opacity with the temperature depends on the {\it amount of disorder} in the amorphous structure and this disorder-induced opacity disappears when the sample is crystallised. When the temperature is high enough, this thermally activated opacity should dominate and determine the spectral shape of the opacity of a given amorphous material. In the TLS model, the thermally activated opacity is due to presence within the amorphous structure of closed rotational-type configurations of atoms or groups of atoms (see \citet{meny2007} for a more detailed description). These may be local defects or impurities. It has been observed for example in various silica samples in which the presence of OH-groups is responsible for the thermally activated opacity \citep{agladze1996,boudet2005}. It may also be groups of atoms participating to the material structure itself. The fact that in our experiments amorphous samples with olivine and pyroxene stoichiometry have different opacity variations suggests that network modifiers such as Mg$^{2+}$ and Ca$^{2+}$ ions do not have well defined positions in the amorphous structure, allowing such closed configurations to exist.\\

\section{Astrophysical Implications}
\label{astro}

\subsection{Relevance of the dust analogues studies}

\noindent Experimental data, even on a limited number of samples, are always relevant because they represent a real behaviour which cannot be ignored. It consists in the present case in the fact that the opacity of amorphous silicate grains varies with temperature, and cannot be described at any temperature with a single spectral index over the whole FIR/mm range. These experimental results are compatible with the anti-correlation between the emission spectral index and the temperature revealed by astrophysical observations from experiments such as PRONAOS, Archeops, Herschel or Planck \citep{dupac2003,desert2008,rodon2010,anderson2010,veneziani2010} and with the submillimetre excess observed by FIRAS/WMAP \citep{paradis2009}. However, at the same time experimental data on interstellar dust analogues must be carefully considered. The comparison of the experimental data with the observations is not straightforward since the astronomical FIR/submm data average the emission from dust grains  that may be coming from different regions of the ISM and/or that may have different sizes, shapes, chemical compositions and temperatures. 

\noindent The studied silicates grains have an uncharacterised size distribution with typical grain sizes of a few tens to a fews hundreds of nm. Interstellar silicate grains, that may have a grain size distribution different from the one of dust analogues, have a typical maximum size of a few hundreds of nm, much smaller than the wavelength in the far infrared. Experiments on silica spherical grains with diameters ranging from 7 to 500 nm show that the measured grain opacity does not depend on the grain size (Coupeaud et al. in preparation). In addition, the absorption part of the refractive index being low, the Rayleigh criteria is fulfilled, and thus the grain size distribution should not influence the resulting dust opacity spectrum. Even though the shape distribution of interstellar grains is unknown, there is no reason for them to be spherical. The polarisation observed in the 9.7 $\mu$m band implies that part of the grains are aligned and thus not spherical \citep{whittet2003}. Interstellar grains could have all possible forms, although they may be rounded by erosion induced by sputtering. The opacities presented in this study are those of a population of submicronic grains having a natural shape distribution resulting from the synthesis method used. They should thus be relatively close to what can be found in the ISM. Interstellar grains may also coagulate to form dust aggregates, in which case the shape, the fluffyness and the properties of the aggregates also need to be taken into account. Experimental studies of dust aggregates are difficult to perform but modelling show that dust aggregation increases the dust emissivity without changing the shape of the emissivity as a function of  wavelength (see Sec.~\ref{mass-kappa}). The studied samples thus represent relevant analogues of interstellar dust in terms of composition but also in terms of size and shape.

\subsection{Classical dust emission models and relevance of the spectral index}
\label{beta-model}
\noindent Because no specific mechanism for absorption and emission was expected in the FIR/mm range, the opacity in that domain was considered to be dominated by the long wavelength wings of the observed mid-infrared vibrational bands. The temperature-independent Lorentz model for bound oscillators is then well appropriate. In the FIR/mm range, far from the MIR resonances, it leads to the simple asymptotic behaviour for the dust opacity indicated Eq.~\ref{kappa2} and for the dust emission for optically thin medium (Eq.~\ref{emission}) to the following expression: 
\begin{equation}\label{greybody}
\mathrm{I(\lambda) = B(\lambda,T) \: M_\mathrm{d} \kappa(\lambda_\mathrm{0}) (\lambda/\lambda_\mathrm{0})^{-\beta} \: d^{-2} \: \Omega^{-1}}
\end{equation}
where $\beta$ = 2. This model fully describes the emission with only two dust parameters: the opacity ${\kappa}$($\lambda_\mathrm{0}$) at a wavelength, ${\lambda}_0$, where the asymptotic behaviour of the Lorentz model is assumed to be fulfilled, and the dust temperature, T, that depends on its radiative environment. The Lorentz model applies for crystals, but it is also used to describe the FIR/mm properties of amorphous solids. The mid-infrared bands, broadened by the dispersion of angles in the amorphous structure, are usually modelled by the sum of closed Lorentz oscillators (up to 33 oscillators are considered in the astrosilicates from \citet{draine1984}). This modelling does not change the asymptotic behaviour in the long wavelength range which is the same as for crystals. The dust emission may also be explained in terms of phonons \citep{henning1997}. Because of the breakdown of the momentum selection rule in amorphous materials, each phonon can be involved in the temperature-independent (or weakly dependent) one-phonon absorption process. This disorder-induced one phonon process gives rise in the FIR/submm to a continuous absorption that completely dominates over multi-phonon absorption processes. As a consequence the dependence of the opacity with the wavelength is expected to follow the one of the phonon density of state g($\lambda$), which follows a simple power law dependence in the Debye approximation: g($\lambda$) $\propto$ $\lambda^{-2}$ in 3-dimensional structures, and g($\lambda$) $\propto$ $\lambda^{-1}$ in 2-dimensional structures \citep{mennella1998}. These two models thus predict that the FIR/mm dust-emission follows at first order the law given by Eq.~\ref{greybody} in which the opacity is temperature-independent and where $\beta$ equals 2, or may vary between 1 and 2. This variation of the spectral index is usually understood in terms of different grain properties: some atomic structures or surface-to-volume ratio which can induce a 2-dimensional behaviour, and/or a distribution of temperatures along the line of sight to justify a decrease of the spectral index to values smaller than 2 \citep{seki1980}.\\

\begin{figure}[!t]
   \includegraphics[scale=.37, angle=90]{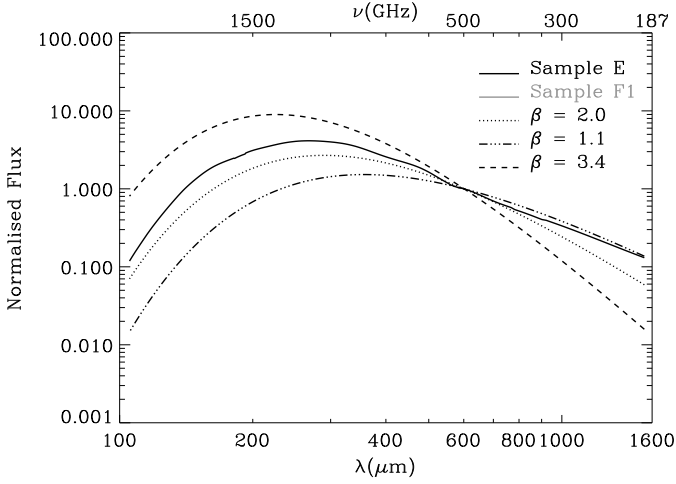}
      \caption{Dust emission simulated using the modified blackbody law (see Eq~\ref{greybody}) for 10K interstellar grains. The solid lines are calculated using the experimental 10K opacity of  the amorphous E (MgSiO$_3$) sample (black line) and the amorphous F1 (Mg$_{2.3}$SiO$_4$) sample (grey line). The dust emission assuming a single spectral index of $\beta$=2.0 (dotted line), $\beta$=3.4 (dotted-dashed line) and $\beta$=1.1 (dashed) are also represented. The curves have been normalized at 600 $\mu$m.}
    \label{astro-f}
\end{figure}

\noindent Within the framework of these models, astronomical observations are thus interpreted with a single, temperature-independent, spectral index value over the whole FIR/mm range. As explained above, such an interpretation relies on the assumption that the knowledge of the dust opacity at a given FIR-wavelength ${\lambda}_0$ (deduced from laboratory measurements at any temperature) and the dust spectral index value allows us to deduce the emission of the dust at any FIR/mm wavelength and at any temperature. However it is clear from the experimental results above that in the amorphous state, the grain opacity, $\kappa$, varies with the temperature and the wavelength. Therefore it is not possible to describe the dust opacity with a single spectral index. It may then be more appropriate to use the local spectral index, $\beta$($\lambda$), define as the slope of the opacity at any wavelength, to take into account the opacity variations with the wavelength. 

\noindent  This has strong implications that question the relevance and the physical meaning of the emissivity spectral index derived from the observations. First of all it means that the use of a single spectral index should not be the rule when modelling the dust emission. Indeed, since the experimental spectral index value of relevant interstellar dust analogues varies with the wavelength, the emissivity spectral index deduced from astronomical data may also vary with the wavelength. As a consequence, observations of an astronomical object at different wavelengths in the FIR/mm range can lead to different values of the emissivity spectral index (and slight variations in the dust temperature). It is interesting to note here that the FIR opacity of the astrosilicate from \citet{draine1984} was modified to account for the high galactic latitude dust emission measured by FIRAS and that, the FIR opacity of the new astrosilicates from \cite{li2001} is wavelength-dependent, the spectral index being equal to 2 below 800 $\mu$m and to 1.6 above. Secondly, it means that the knowledge of the opacity at a single FIR wavelength and a single spectral index, either expected from simple theory ($\beta$=2) or measured in a given limited wavelength range, is not sufficient to derive accurately the emission of the dust on the whole FIR/mm domain. These two points are well-illustrated by Table~\ref{beta-t} and Fig.~\ref{astro-f} that shows the simulated emission of two amorphous grain populations of mean composition close to MgSiO$_3$ and Mg$_2$SiO$_4$. For $\lambda \le$ 600 $\mu$m, the derived spectral index is $\beta \sim$ 2 whereas it is $\sim$ 4 at $\lambda \ge$ 600 $\mu$m for the grains with an olivine stoichiometry and, respectively,  $\beta \sim$ 2.5 and $\sim$ 1.1 for the grains with an enstatite stoichoimetry.

\begin{figure}[!t]
   \includegraphics[scale=.37, angle=90]{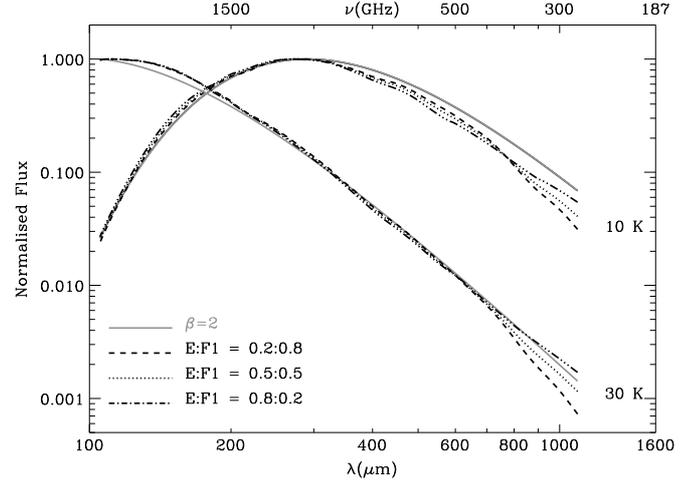}
      \caption{Dust emission simulated using the modified blackbody law (see Eq~\ref{greybody}) for a {mixture of amorphous grains with enstatite and olivine stoichiometry}, for two grains temperatures: 10 and 30 K. The solid grey lines are calculated using a single spectral index, $\beta$=2. The amorphous E sample with an pyroxene stoichiometry and the F1 sample with an olivine stoichiometry} have been used for the modelling with the following respective abundances: 0.2:0.8 (dashed line), 0.5:0.5 (dotted line) and 0.8:0.2 (dotted-dashed line). The curves have been normalized to unity.
    \label{mix-sil}
\end{figure}

\noindent However, the fact that the dust emission in certain astronomical environments are well-modelled with a single temperature and emissivity spectral index value \citep{planck2011u} is not in contradiction with the experimental results presented in this study. In this particular case, a population of amorphous grains with olivine and pyroxene stoichiometry may produce a resulting  emissivity spectral index constant over the whole spectral range (Fig.~\ref{mix-sil}). Actually, such a case illustrates well the difference between the experimental spectral index and the observational emissivity spectral index that must be carefully used to describe the dust as it averages contributions from various dust composition. It also illustrates that the variation of the dust spectral index towards various lines of sight may trace changes in the dust composition.\\

\subsection{Dust mass and opacity}
\label{mass-kappa}
\noindent These experimental results have also consequences on the use of methods based on flux ratios to determine the dust temperature and spectral index that are commonly used in absence of spectroscopic data. \citet{shetty2009b} investigated the effect of line of sight temperature variation and noise on the accuracy of the derived parameters such as the dust temperature and spectral index or the dust mass. However, another important source of error comes from the fact that the flux ratios method is based on a single value for the spectral index and from uncertainties on the MAC itself. From the experimental results presented here it is also possible to investigate this effect. Flux ratio are proportional to the ratio of the blackbody emissions at the dust temperature and to the ratio of the opacities, at the two wavelengths. The error may be important if the fluxes are observed at wavelengths that are very different and situated on both sides of $\sim$ 600-800 $\mu$m, where the local spectral index of dust analogues changes. As an example the ratio of the experimental MAC at 450 and 1200 $\mu$m for the E sample (pyroxene stoichiometry) is  $\kappa_\mathrm{450}$ / $\kappa_\mathrm{1200} \sim$ 4.8 and 3.7 at 10K  and 30 K, respectively, whereas, using the asymptotic expression for the dust MAC (Eq~\ref{kappa2}), $\kappa_\mathrm{450}$ / $\kappa_\mathrm{1200} \sim$ 7.1 for $\beta$ = 2 and  $\kappa_\mathrm{450}$ / $\kappa_\mathrm{1200} \sim$ 11.6 for $\beta$ = 2.5 (value that reproduces the MAC of the E sample at $\lambda \le$ 600 $\mu$m). Hence, the flux ratio is wrong by a factor of 1.5-2 if one adopt $\beta$=2 and by a factor of 2.4-3 if one adopt $\beta$=2.5. If the flux ratio is calculated from observations at 450 and 850 $\mu$m, the experimental ratio $\kappa_\mathrm{450}$ / $\kappa_\mathrm{850}$ is equal to 3.6 and 2.6 for the E sample at 10 and 30 K and  $\kappa_\mathrm{450}$ / $\kappa_\mathrm{850}$ = 3.6 and 4.9 using the asymptotic expression for the dust MAC (Eq~\ref{kappa2}) and $\beta$ =2 and 2.5 respectively. Hence the flux ratio is wrong by a factor of 1-1.4 and 1.4-1.9 if one adopt $\beta$ equals 2 and 2.5, respectively. The error on the flux ratios is thus larger when data at 450 and 1200 $\mu$m are used. For the same reasons it has some consequences on the determination of masses from FIR/submm data. At 10 K, the experimental MAC of the E sample is $\kappa_{1200{\mu}m} \sim$ 0.45 whereas the value extrapolated using the asymptotic expression for $\kappa$ is 0.17, at 850 $\mu$m $\kappa_{850{\mu}m} \sim$ 0.7 whereas the extrapolated value is  0.45. The error on the mass is thus of a factor 3 and 2 at 1200 and 850 $\mu$m, respectively.

\noindent The intrinsic MAC for a given material/grain distribution as measured in laboratory increases as the temperature increases. However, astrophysical observations reveal spatial variations of the MAC, and show that the dust emissivity at long wavelength seems to increase from the diffuse medium to dense and cold regions by a factor of 3 to 4 \citep{stepnik2003,juvela2011,planck2011u}. Such results seem at first sight to contradict the temperature dependence of the MAC observed experimentally as a general behaviour in all amorphous samples, but these two experimental and observational facts are not necessarily contradictory. To date, the physical interpretations associated with these two facts appear different and complementary. Experimentally, we measure the variation with the temperature of the MAC of a given population of specific silicate grains and as a general fact the MAC appears correlated with the temperature. Such behaviour is due to the intrinsic physical properties of the amorphous state. The disorder of the amorphous state induces some additional low energy broad band to the MAC that have to be taken into account and we cannot consider anymore that the FIR/submm MAC in amorphous silicates is only dominated by the long wavelength wings of the mid infrared bands. As discuss above it can be understood in terms of the TLS model. In contrast, the observational spatial variations of the interstellar dust MAC are analysed in terms of a spatial evolution of the grain population: aggregation, fragmentation, variations of both the size distribution and the fluffiness or fractal dimension of the grains due to the aggregation processes \cite{stepnik2003}. The increase of emissivity observed in the core of dense clouds where the dust temperature is low probably reflects a variation of the grain population, through a change in the composition and the morphology of the grains due to the aggregation/fragmentation processes and not a variation with the temperature of the intrinsic properties of the grain material, as it is observed experimentally. Model calculations with the discrete-dipole approximation (DDA) can explain this enhancement with the coagulation of dust particles into aggregates. In the FIR/mm range, such coagulation increases the amplitude of the observed emissivity of the dust population without changing its spectral profile (i.e. spectral index)  \citep{ossenkopf1994,kohler2011}. However, DDA and similar calculations require the knowledge of the intrinsic optical properties of the material that constitutes the grains (complex dielectric constant or refractive index), and these intrinsic optical properties used at the basis of the modelling have to take into account their temperature-dependences.\\

\subsection{Interpretation of $\beta$-T diagrams}
\label{beta-T}

\noindent Results from experimental work from this study but also from \citet{agladze1996,mennella1998,boudet2005} have implications on the derivation and interpretation of $\beta$-T diagrams that are used to investigate the dust properties in various astrophyscial environments. In particular it shows that it is difficult, if not impossible, to compare T-$\beta$ diagrams if they rely on observations with different spectral coverages. If the diagrams are derived from observations whose spectral coverage does not overlap, then their comparison is  meaningless because it is impossible to conclude anything on the dust properties since the same dust population may have very different emissivity spectral index in different domains (see Figs.~\ref{astro-f} and \ref{mix-sil}). However this is almost never the case since data at short wavelengths ($\sim$ 100 $\mu$m) are necessary to constrain the dust temperature. If the spectral coverage overlaps, as it is the case for example for Herschel and combined Herschel/Planck data or Herschel/ground based data, then the derived single emissivity spectral index, which is averaged over the wavelength, is thus different for the two spectral coverages. The more the spectral coverages overlap, the more the derived spectral indexes from the two dataset are comparable. This should be kept in mind, in addition to the effect of observational noise on the derivation of the dust temperature and emission spectral index \citep{shetty2009b} and to the effect of temperature variations along the line of sight \citep{juvela2011}, when interpreting $\beta$-T diagrams. Nevertheless, there is no doubt from the experiments presented in this study, but also from previous experimental studies from \citet{agladze1996,mennella1998,boudet2005}, that an anti-correlation between the silicate grain temperature and the spectral index $\beta$ intrinsic to the grain material does exist (Table~\ref{beta-t}). Hence, even though observational bias such as noise and temperature distribution certainly complicate the data analysis and reduce the accuracy of the derived parameters, the $\beta$-T anticorrelation, intrinsic to the grain material, exists and certainly contributes to the observed  T-$\beta$ anticorrelation. 

\noindent Indeed, the experimental MAC variations with temperature are in qualitative agreement with astrophysical observations ($\beta$-T anti-correlation and millimetre excess). However the experimental $\beta$-T anti-correlation occurs in a wider temperature range than the one observed in astrophysics: except for the samples with an enstatite stoichiometry, we do not observe significant MAC variation within the 10-30 K range. This may have several causes. First of all, for all the reasons discussed above, the astronomical $\beta$-T anti-correlation can be only partly attributed to variation of grains properties with temperature. Secondly, as suggested by the experimental results, in agreement with the TLS model predictions, the MAC variations are due to disorder in the amorphous state (see Sect.\ref{discussion}). Thus, astronomical silicates may present more (or less) pronounced variation of their MAC with T and $\lambda$ than the dust analogues studied here, according to their micro-structure and composition. The comparison and modelling of further astrophysical and experimental observations with the same physical model, such as the TLS model, should be useful to discriminate between these points.\\

\subsection{Perspectives}
\label{persc}
\noindent From a theoretical point of view, the emission law formulated in Eq.~\ref{emission} for diffuse interstellar dust should be relativized since it cannot take into account the dependence of the dust MAC with the temperature. \citet{meny2007} proposed an alternative physical model for the dust emission, the TLS model, based on specific properties of the amorphous state. We show that in the disordered structure of the amorphous grains, the charge disorder and the density of states of low energy levels (which can be seen as a distribution of closed rotational-type configurations of atoms or groups of atoms in the amorphous structure) lead to absorption processes that can dominate, in the FIR/mm wavelength range, over the absorption of the mid-infrared resonance wings. This model, which is in good agreement with astrophysical observations \citep{paradis2011} and laboratory-experiments (see Sect.~\ref{discussion} and Meny et al. in preparation), is now being improved (Gromov et al. in preparation).\\

\noindent  We have presented here a study of Mg-rich amorphous silicate grains. However it must be kept in mind that interstellar grains are probably more complex than the studied samples.  In particular they may contain iron in the form of oxide or metallic inclusions or within the silicate network. They may also be coated with ices in dense and cold clouds or with carbonaceous matter. Grains of carbonaceous matter may also be big enough to radiate at thermal equilibrium in the FIR/submm range. These material will certainly influence the shape of the dust emission at long wavelength. However, the optical properties of ices and carbonaceous matter in the FIR/submm at low temperature are poorly known. \citet{mennella1998} studied two amorphous carbon samples in the range 20-2000 $\mu$m, from 300 to 24 K.  As for the silicate grains they found that the MAC decreases when the grain temperature decreases and that its spectral shape is slightly modified. These experimental results must be taken into account for the interpretation of FIR/submm/mm data such as the Herschel and Planck data.  Carbonaceous dust is not as well constrained as the silicate dust and further experimental work have to be done to characterise the FIR/submm spectroscopic properties at low temperature of this dust component but also of ices, oxides and Fe-rich silicate dust.

\section{Conclusion}

In this study we report the characterisation of the far infrared and submillimetre optical properties of various interstellar dust analogues as a function of  grain temperature. The studied samples are amorphous and crystallised Mg- and Ca-rich nanograins of mean composition characteristic of forsterite, enstatite and diopside. For all the {\it amorphous} samples, independently of their composition, we observe that: 
\begin{list}{\arabic{qcounter}.~}{\usecounter{qcounter}}\addtocounter{qcounter}{0}
\item The FIR/submm mass absorption efficiency (MAC), or opacity, decreases when the grain temperature decreases. 
\item The spectral shape of the MAC curve does not follow a simple asymptotic behaviour with the wavelength so that it cannot be described correctly with a single spectral index. 
\item At a given wavelength, the locally defined spectral index (i.e. the slope of the MAC curve) increases when the dust temperature decreases.
\end{list}
\noindent These variations of the grain MAC, that are not observed for crystallised samples, are due to the amorphous nature of the samples. However, the detailed spectral shape of the MAC depends on the cations/Si ratio of the amorphous materials and is different for amorphous materials having olivine and pyroxene stoichiometry. At short wavelength ($\lambda \le 500/700 \mu$m), the spectral index $\beta$ is close to the value of 2 (1.6 $ \le \beta \le$ 2.2) for all grain temperature and composition whereas at longer wavelengths, $\beta \le$ 2 for the pyroxene stoichiometric samples ($\beta$ = 0.9 for the Mg$_{0.95}$SiO$_3$ sample at 10 K above 800 $\mu$m) and $\beta \ge$ 2 for the olivine stoichiometric samples ($\beta$ = 3.6 for the Mg$_{2.3}$SiO$_4$ sample at 10 K above 700 $\mu$m). 

\noindent These experimental results have important implications for the interpretation of astronomical observations. They emphasise the fact that adopting a power law characterised by a single spectral index independent of the wavelength and the grain temperature for the dust MAC is not appropriate and may induce significant errors on the derived physical parameters such as the dust temperature and mass. The study of interstellar dust should be based either on the MAC as a function of the wavelength and temperature rather than on the spectral index. Alternative models such as the TLS model that reproduces the variation of the grain optical properties with the temperature and wavelength may also be considered. Then, new information about the dust, its physical and chemical environment and its evolution may be obtained.

\begin{acknowledgements}
The authors would like to thank Nicolas Ratel-Ramond and Yannick Coppel for the XRD and NMR measurements and Mike Toplis for fruitful discussions. This paper has benefit from stimulating and critical discussions held within the "Cold Dust" project collaboration and at the "Herschel and the Characteristics of Dust in Galaxies" workshop (28 Feb - 04 March 2001, Leiden). The authors would like to thank the referee, Th. Henning, for his critical review of this paper. We acknowledge SOLEIL for provision of synchrotron radiation facilities. Support by the French National Agency (ANR-07-BLAN-0364-01, "Cold Dust") and from the european program FEDER is acknowledged.
\end{acknowledgements}

\bibliographystyle{aa}
\bibliography{bib.bib}

\end{document}